\documentclass[reprint,prl,twocolumn,showpacs,superscriptaddress,aps,longbibliography]{revtex4-2}

\usepackage[colorlinks=true, citecolor=blue, linkcolor=blue]{hyperref} 
\usepackage{graphicx}
\usepackage{bm}
\usepackage{amsmath}
\usepackage{amssymb}
\usepackage{svg}

\DeclareGraphicsExtensions{.png,.jpg,.eps,.svg}
\usepackage{xcolor}

\newcommand{\beq}{\begin{equation}}
\newcommand{\eeq}{\end{equation}}

\usepackage{mleftright} 

\newcommand{\figref}[1]{\mbox{Fig.~\ref{#1}}}

\renewcommand{\eqref}[1]{\mbox{Eq.~(\ref{#1})}}

\newcommand{\figpanel}[2]{Fig.~\hyperref[#1]{\ref*{#1}(#2)}}
\newcommand{\figpanels}[3]{Fig.~\hyperref[#1]{\ref*{#1}(#2)-(#3)}}
\newcommand{\figpanelNoPrefix}[2]{\hyperref[#1]{\ref*{#1}(#2)}}

\newcommand{\ket}[1]{|#1 \rangle}
\newcommand{\bra}[1]{\langle #1|}
\newcommand{\braket}[2]{\langle #1|#2 \rangle}

\usepackage{siunitx}

\begin{document}

\author{Walter Rieck}
\affiliation{Department of Microtechnology and Nanoscience, Chalmers University of Technology, 41296 Gothenburg, Sweden}


\author{Anton Frisk Kockum}
\email{anton.frisk.kockum@chalmers.se}
\affiliation{Department of Microtechnology and Nanoscience, Chalmers University of Technology, 41296 Gothenburg, Sweden}

\author{Guangze Chen}
\email{guangze@chalmers.se}
\affiliation{Department of Microtechnology and Nanoscience, Chalmers University of Technology, 41296 Gothenburg, Sweden}

\title{Doublon bound states in the continuum through giant atoms}

\begin{abstract}

Bound states in the continuum (BICs) are spatially localized modes embedded in the spectrum of extended states, typically stabilized by symmetry or interference. While extensively studied in single-particle and linear systems, the many-body regime of BICs remains largely unexplored. Here, we demonstrate that giant atoms, quantum emitters coupled nonlocally to structured waveguides, can host robust doublon BICs, i.e., two-photon bound states stabilized by destructive interference and interactions. We first analyze a driven two-photon emission process and show how doublon BICs arise and mediate decoherence-free interaction between distant atoms. We then demonstrate that these many-body BICs also emerge under natural, undriven dynamics via a virtual two-photon emission process in three-level giant atoms. Our results reveal an interference-based mechanism for stabilizing many-body localization in open quantum systems, with potential applications in quantum simulation, non-ergodic dynamics, and protected quantum information processing.

\end{abstract}

\date{\today}

\maketitle


\textit{Introduction.---}Bound states in the continuum (BICs) are spatially localized eigenstates that exist within the continuous spectrum of extended, radiative modes~\cite{vonNeumann1993, PhysRevA.11.446}. They can arise through mechanisms such as symmetry protection~\cite{PhysRevLett.107.183901}, separability of variables~\cite{Robnik1986}, or destructive interference between leakage channels~\cite{PhysRevA.32.3231, PhysRevLett.100.183902, PhysRevLett.113.037401}. BICs have been extensively studied in photonic and acoustic systems~\cite{Hsu2016, Koshelev2020, Joseph2021, Azzam2020, PhysRevLett.128.084301, Tong2020}, where their unique combination of spatial confinement and spectral embedding gives rise to exceptionally long lifetimes and sharp resonances. Crucially, their in-band nature enables efficient coupling to external radiation or nearby modes, while their bounded character ensures strong field localization. These properties make BICs ideal platforms for many applications, including nonlinear optical enhancement~\cite{PhysRevLett.121.033903, PhysRevLett.123.253901, Koshelev2019, PhysRevResearch.1.023016, Anthur2020, Zograf2022}, coherent light generation~\cite{Kodigala2017, Ha2018, Wu2020, Wang2021highly}, and ultrasensitive sensing~\cite{Yesilkoy2019, Romano2019, Tseng2020, Altug2022, Chen2020}. 

Recently, there has been growing interest in extending the concept of BICs to the many-body regime, where they could offer new mechanisms for suppressing thermalization~\cite{sugimoto2023manybodyboundstatescontinuum}. However, this regime remains largely unexplored. Existing studies typically rely on impurity-induced localization~\cite{PhysRevLett.109.116405, PhysRevResearch.4.023194, sugimoto2023manybodyboundstatescontinuum}, which requires engineered on-site potentials and is therefore experimentally challenging, or on boundary effects~\cite{Sun2024}, where the resulting BICs are tied to the system’s boundary and offer limited flexibility for applications. This motivates the search for new platforms and mechanisms to realize many-body BICs in a more controllable manner.

Giant atoms (GAs)---quantum emitters coupled to a waveguide at multiple, spatially separated points, have emerged in the past decade as a versatile platform for exploring nonlocal and interference-driven light–matter interactions~\cite{Kockum2021}. As shown both theoretically~\cite{Kockum2014, Guo2017, Kockum2018, Gonzalez-Tudela2019, Guo2020, Guimond2020, Ask2020, Cilluffo2020, Wang2021, Du2021, Soro2022, Wang2022, Du2022, Du2022a, Terradas-Brianso2022, Soro2023, Du2023, Ingelsten2024, Leonforte2024, Wang2024, Roccati2024, Gong2024, Du2025, Du2025a} and experimentally for several systems~\cite{Gustafsson2014, Manenti2017, Satzinger2018, Bienfait2019, Andersson2019, Kannan2020, Bienfait2020, Andersson2020, Vadiraj2021, Wang2022a, Joshi2023, Hu2024, Almanakly2025, Jouanny2025}, the nonlocal coupling of GAs results in interference effects that alter spontaneous emission and give rise to phenomena beyond conventional quantum optics. A striking example is the emergence of decoherence-free interaction (DFI)~\cite{Kockum2018, Kannan2020}, where multiple GAs interact coherently through a waveguide without radiative losses into the waveguide. Notably, the DFI can be understood as a manifestation of BICs~\cite{Guo2020, Terradas-Brianso2022, PhysRevA.102.033706, Ingelsten2024}, where destructive interference between emission pathways cancels radiation into the waveguide, forming long-lived collective modes. This connection highlights the potential of GAs to realize interference-based many-body BICs. However, studies of BICs in giant-atom systems have so far been restricted to the single-excitation regime, leaving the many-body dynamics and localization mechanisms unexplored.

In this Letter, we propose a mechanism to realize many-body BICs through destructive interference in giant-atom systems, focusing on the emergence of doublon BICs---two-photon bound states that remain localized despite lying within the radiation continuum of a structured waveguide. We begin by analyzing a driven two-photon emission process for a single two-level GA, demonstrating how these states arise from the interplay between interactions and interference, which suppresses decay into the waveguide. We then show that doublon BICs can mediate DFI between two GAs, reinforcing their physical significance. Going beyond this illustrative setup, we demonstrate that doublon BICs and their associated DFI can also emerge under undriven dynamics with realistic coupling configurations for three-level GAs. We finally discuss how our proposal can be realized on several experimental platforms and its potential applications in generating and distributing many-body entangled states.


\begin{figure}
\center
\includegraphics[width=\linewidth]{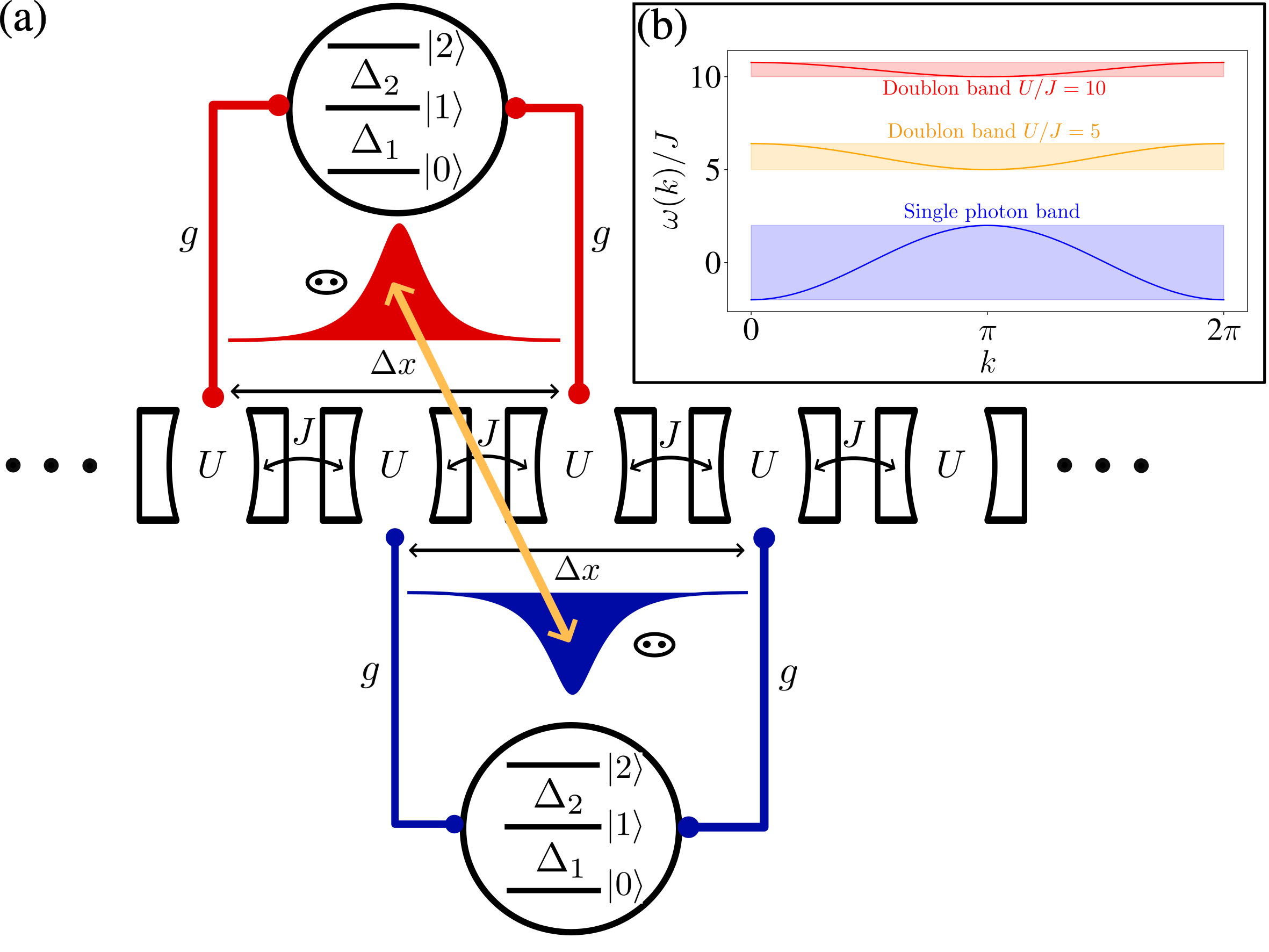}
\caption{Overview of the system.
(a) Schematic of two three-level GAs coupled to a structured waveguide with nearest-neighbor hopping rate $J$ and on-site two-photon interaction $U$. Each atom has level spacings detuned by $\Delta_{1,2}$ from the band center and is coupled to the waveguide at two points, spatially separated with distance $\Delta x=2$, forming a braided configuration. The coupling strength at each connection point is $g$. For appropriate level spacings, doublon bound states in the continuum (BICs) emerge via interference between emission pathways. The overlap of these doublon BICs between the two atoms mediates a decoherence-free interaction (orange arrow).
(b) Band structure of the nonlinear waveguide, for two different values of U.
}
\label{fig:setup}
\end{figure}

\textit{Model.---}We consider a system of multiple GAs coupled to a structured waveguide; see \figpanel{fig:setup}{a}. The total Hamiltonian for this system is
\beq \label{eq:Htot}
H = H_A + H_B + H_{\rm int},
\eeq
where $H_A$ and $H_B$ are the bare Hamiltonians for the atoms and the bath (waveguide), respectively, and $H_{\rm int}$ describes their interaction.

The bath we consider is a one-dimensional array of $N$ coupled cavities with nearest-neighbor hopping $J$ and on-site Kerr nonlinearity $U$. Without loss of generality, we work in a frame rotating at $\omega_c$, the center of the bath band. The bare bath Hamiltonian in this rotating frame is thus
\beq \label{eq:HB}
H_B = \sum_{n = 1}^N \frac{U}{2} a_n^\dag a_n^\dag a_n a_n - \sum_{n = 1}^{N-1} J \mleft( a_n^\dag a_{n+1} + \text{H.c.} \mright),
\eeq
where $a_n$ ($a_n^{\dag}$) is the photon annihilation (creation) operator for cavity $n$ and H.c.~denotes Hermitian conjugate. 

The Kerr nonlinearity $U$ supports the formation of doublons---two-photon bound states where the photons tend to stay near each other~\cite{Winkler2006, PhysRevA.76.023607, PhysRevLett.109.116405, PhysRevLett.124.213601, PhysRevA.95.053866}. A doublon with momentum $k$ has the wavefunction~\cite{PhysRevA.76.023607, PhysRevLett.124.213601}
\beq
\ket{\psi_D(k)} = \frac{1}{\sqrt{2}} \sum_{m=1}^N\sum_{n=1}^Ne^{ik\frac{(m+n)}{2}}e^{-\frac{|m-n|}{\varsigma(k)}}a^\dag_ma^\dag_n\ket{\text{vac}}
\eeq
where $\varsigma(k)$ is the momentum-dependent localization length and $\ket{\text{vac}}$ is the vacuum state. These doublons can propagate through the waveguide and have the dispersion relation~\cite{PhysRevA.76.023607,PhysRevLett.124.213601}
\beq \label{eq:DoDR}
\omega_{D}(k) = \operatorname{sign}(U) \sqrt{U^2 + 16 J^2 \cos^2{\mleft( \frac{k}{2} \mright)}}.
\eeq
Notably, when $U\geq 4J$, the doublon band becomes energetically separated from the continuum of two unbound photons; see \figpanel{fig:setup}{b}. This separation allows us to isolate doublon-mediated dynamics by tuning the atomic level spacings to match the doublon dispersion.

Since our study involves multi-photon processes, we model each GA as a three-level system. For simplicity, we assume all atoms have an identical ladder-type level structure, with $\omega_{1}$ ($\omega_2$) the level spacing between the $\ket{0}$ ($\ket{1}$) and $\ket{1}$ ($\ket{2}$) levels of the atom. The bare Hamiltonian for $N_a$ GAs is then, in the frame rotating at $\omega_c$,
\beq \label{eq:HA}
H_A = \sum_{n=1}^{N_a} \mleft[ {\mleft(\Delta_2 + \Delta_1\mright)} \frac{\sigma_n^{(2), +} \sigma_n^{(2), -}}{2} + \Delta_1 \sigma_n^{(1), +} \sigma_n^{(1), -} \mright],
\eeq
where $\Delta_j = \omega_j - \omega_c$ and the ladder operators act as
\beq
\sigma_n^{(j), +} \ket{j-1} = \sqrt{j} \ket{j}, \quad
\sigma_n^{(j), -} \ket{j} = \sqrt{j} \ket{j-1}.
\eeq


\begin{figure}[t!]
\center
\includegraphics[width=\linewidth]{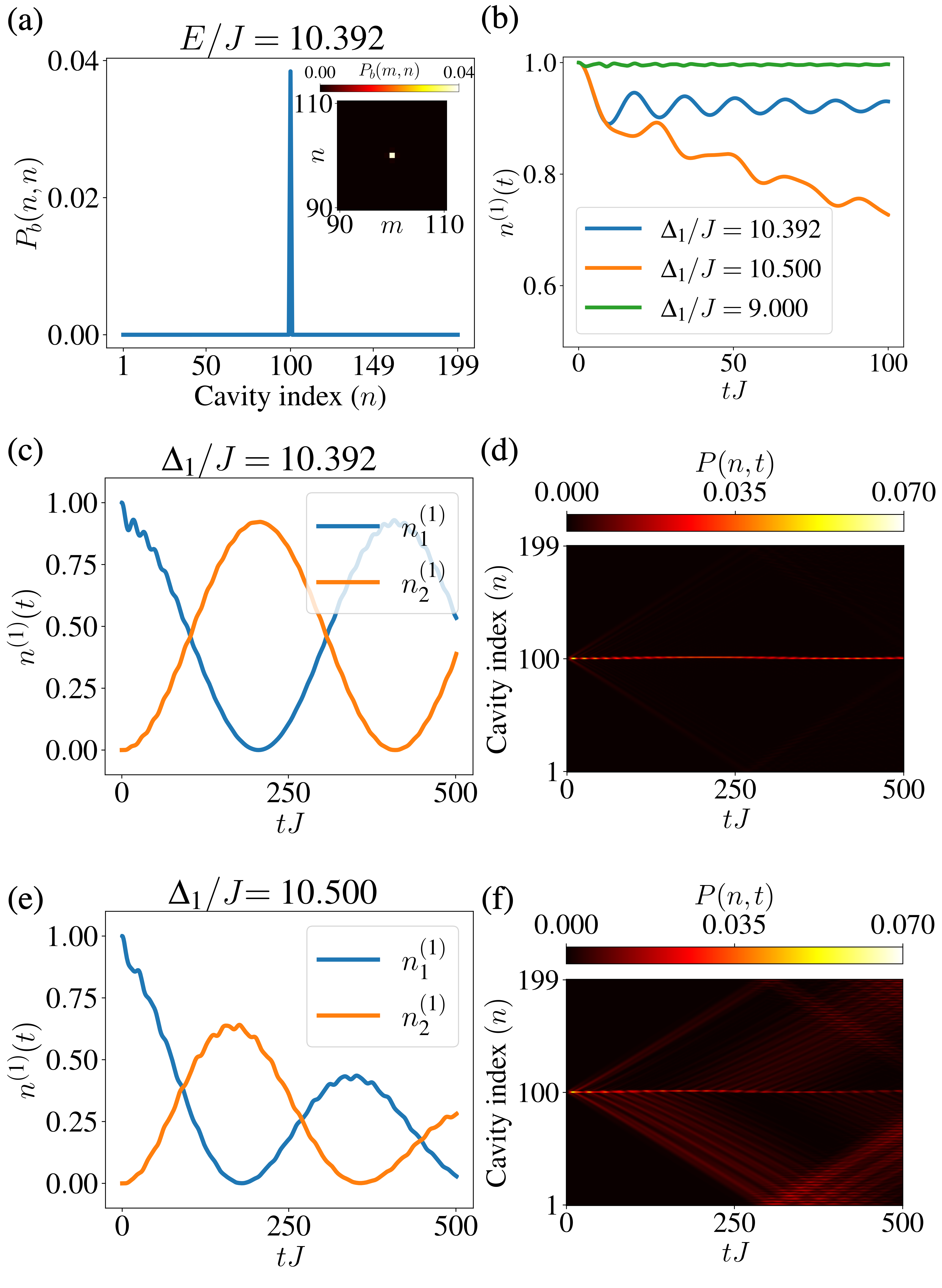}
\caption{Doublon BIC and DFI between GAs, for two-photon interaction [\eqref{eq:HDC}]. Parameters: $U = 10J$, $g = 0.04 J$, and $\Delta x = 2$.
(a) The doublon BIC for a single GA with the DF frequency $\Delta_1 = 10.392J$. The plot shows the bound state's overlap $P_b(n,n)$ with the two-photon states $\ket{0,nn}$, while the inset also shows the off-diagonal elements $P_b(m,n)$.
(b) Time evolution of the population $n^{(1)}(t)$ of the GA's $\ket{1}$ level for three values of $\Delta_1$: at the DF frequency (blue), inside the doublon band (orange), and outside the doublon band (green).
(c) Population dynamics $n^{(1)}_{1,2}(t)$ of two braided GAs at the DF frequency.
(d) Heatmap of the doublon population $P(n,t)=\bra{\psi(t)}a^\dag_na^\dag_na_na_n\ket{\psi(t)}$ in the waveguide. 
(e, f) Same as in (c, d), but for a non-DF frequency inside the doublon band. 
}
\label{fig2}
\end{figure}

\textit{Direct two-photon emission.---}The key ingredient of GAs is the multiple-point coupling: each GA $n$ is coupled to the waveguide at $p_n$ different cavities with strength $g$. To illustrate how this multi-point coupling can result in a doublon BIC, we first consider a direct two-photon emission process described by the interaction Hamiltonian
\beq \label{eq:HDC}
H_{\rm int} = g \sum_{n=1}^{N_a} \sum_{j=1}^{p_n} \mleft[\sigma_n^{(1), +} a_{x_{nj}} a_{x_{nj}} + \text{H.c.} \mright] ,
\eeq
where $x_{nj}$ is the $j$th coupling point of atom $n$. Although we here mostly use this type of nonlinear coupling for illustrative purposes, it can be engineered experimentally using external drives~\cite{PhysRevA.110.053711}. If the two-photon operator in \eqref{eq:HDC} is replaced with a single-photon operator, it is known that single-photon BICs emerge due to destructive interference between multiple coupling points~\cite{Kockum2014}. In our case, since two photons are created (or annihilated) at the same site, the resulting state has significant overlap with doublon modes. Thus, we expect a doublon BIC if the atomic level spacing $\Delta_1$ lies within the doublon band.

To demonstrate this explicitly, we simulate a system of $N=199$ cavities, where a single GA couples to sites $x_{11}=99$ and $x_{12}=101$, i.e., a separation $\Delta x=2$. A doublon BIC arises due to destructive interference between emission paths~\cite{Kockum2014}, governed by the condition $1+\exp(ik\Delta x)=0$, which is satisfied with
\beq
k_\text{DF}=\frac{(2n+1)\pi}{\Delta x},\quad n\in\mathbb{N}.
\eeq
For $\Delta x=2$, this yields $k_\text{DF}=\pm\pi/2$; the corresponding doublon frequency is $\omega_\text{DF}=\omega_D(k_\text{DF})$. Thus, setting $\Delta_1=\omega_\text{DF}$ enables the formation of a doublon BIC, which is shown in \figpanel{fig2}{a} for $U = 10 J$ and $g = 0.04 J$ (chosen to remain in the Markovian regime).

We identify the BIC state $\ket{\psi_b}$ using the inverse participation ratio~\cite{wegner1980inverse,footnote1}. Its overlap with the two-photon basis state $\ket{0,mn}=a^\dag_ma^\dag_n\ket{\text{vac}}$, denoted $P_b(m,n)=\mleft|\braket{0,mn}{\psi_b}\mright|$, confirms that it is a doublon BIC: the overlap becomes exponentially small when $m\neq n$ [inset of \figpanel{fig2}{a}], a typical characteristic of a doublon; additionally, the wavefunction is localized in space, demonstrating its bound nature. 

In \figpanel{fig2}{b}, we plot the atom's $\ket{1}$-level population $n^{(1)}(t) = \bra{\psi(t)} \sigma^{(1),+}\sigma^{(1),-} \ket{\psi(t)}$ for three different atomic frequencies, starting from $\ket{\psi (t=0)} = \sigma^{(1),+}\ket{\text{vac}}$. When $\Delta_1=\omega_\text{DF}$ (blue curve), this population stabilizes at a high value, typical of a BIC. In contrast, for $\Delta_1$ within the doublon band (orange curve), the atom population displays exponential decay due to coupling to radiative modes instead of the doublon BIC. For $\Delta_1$ outside the doublon band (green curve), the atomic population shows decoherence-free behavior due to the formation of a bound state outside the continuum (BOC).

A key distinction between BICs and conventional BOCs is that BICs reside within the band, enabling significantly faster DFI between multiple GAs. To illustrate this, we add a second GA coupled to cavities 100 and 102, forming the braided configuration shown in \figpanel{fig:setup}{a}. This configuration is known to be essential for enabling overlap between BICs associated with different atoms, a prerequisite for DFI~\cite{Kockum2018, Ingelsten2024}. As shown in \figpanel{fig2}{c}, starting from $\ket{\psi (t=0)} = \sigma_1^{(1),+}\ket{\text{vac}}$, this setup leads to coherent oscillations between the two atoms, except for a leakage at early times due to non-Markovian effects~\cite{Soro2023}. The corresponding doublon population in the waveguide $P(n,t)=\bra{\psi(t)}a^\dag_na^\dag_na_na_n\ket{\psi(t)}$ is visualized in \figpanel{fig2}{d}, revealing no doublon leakage outside. In contrast, a slight detuning from $\omega_\text{DF}$ not only suppresses the formation of BIC, but also allows partial population of the extended doublon modes. As a consequence, the DFI is disrupted [\figpanel{fig2}{e}] and doublon leakage into the waveguide becomes prominent [\figpanel{fig2}{f}].




\begin{figure}[t!]
    \centering
    \includegraphics[width=\linewidth]{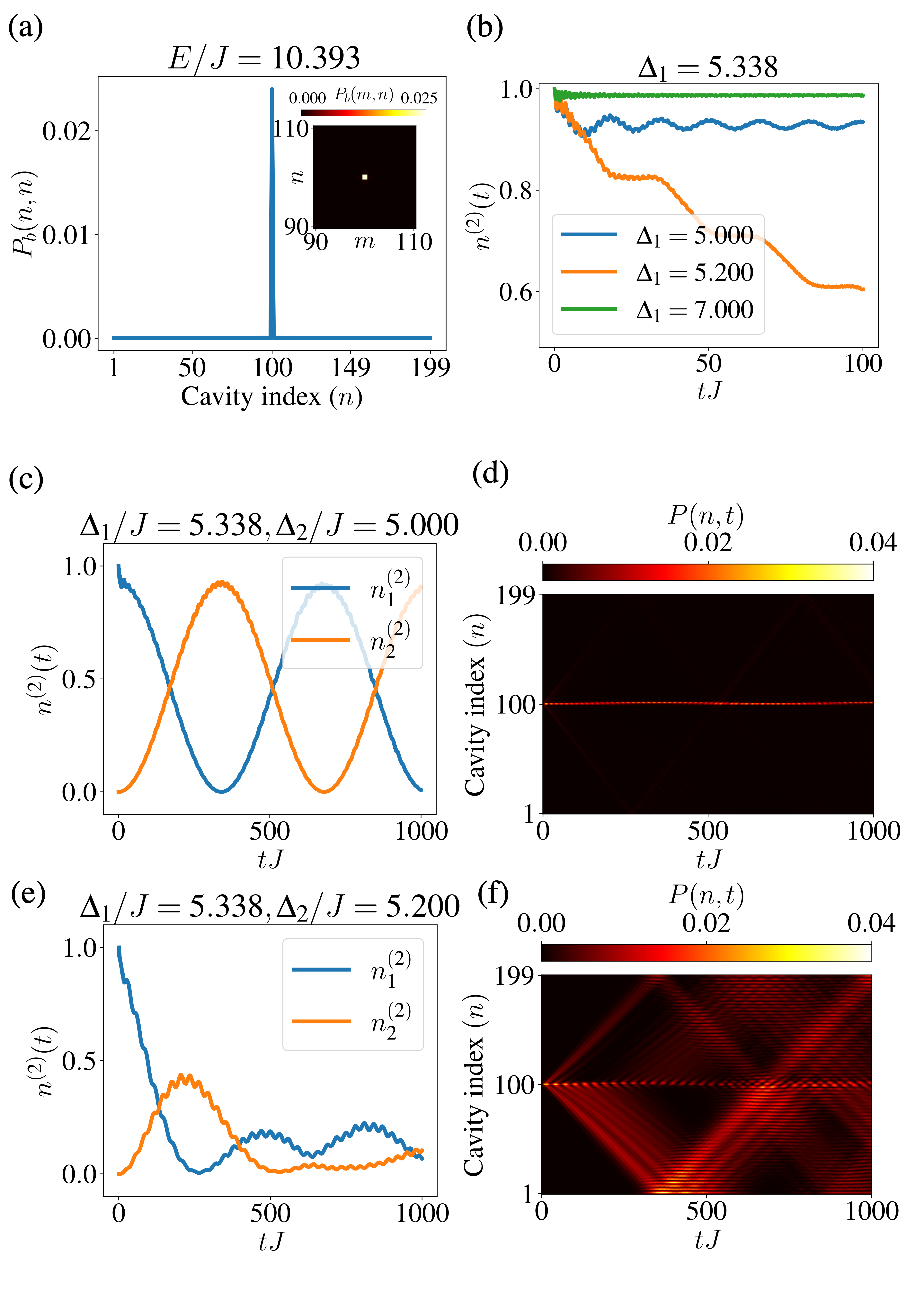}
    \caption{Doublon BIC and DFI between GAs, for single-photon coupling [\eqref{eq:Hnat}]. Parameters: $U = 10J$, $g = 0.25 J$, and $\Delta x = 2$.
    (a) The doublon BIC for a single GA with the DF condition $\Delta_1 = 5.338J$ and $\Delta_2 = 5J$. The plot shows the diagonal overlap $P_b(n,n)$, indicating strong localization in real space. The inset shows the full two-photon amplitude $P_b(m,n)$, with vanishing off-diagonal elements confirming the doublon nature.    
    (b) Time evolution of the population $n^{(2)}(t)$ of the GA's $\ket{2}$ level for fixed $\Delta_1 = 5.338J$ and varying $\Delta_2$: at the DF condition (blue), inside the doublon band (orange), and outside the doublon band (green).
    (c) Population dynamics $n^{(1)}_{1,2}(t)$ of two braided GAs at the DF frequency.
    (d) Heatmap of the doublon population $P(n,t)=\bra{\psi(t)}a^\dag_na^\dag_na_na_n\ket{\psi(t)}$ in the waveguide.
    (e, f) Same as in (c, d), but for a non-DF frequency inside the doublon band.}
    \label{fig3}
\end{figure}

\textit{Single-photon coupling.---}Above, we demonstrated the emergence of doublon BICs and the associated DFI between GAs coupled to a structured waveguide, using the direct two-photon coupling in~\eqref{eq:HDC}. While that example is illustrative and conceptually clear, the two-photon coupling that it relies on generally requires external drives. We now turn to a more practical, and theoretically more intriguing, scenario based on common single-photon coupling with three-level GAs:
\beq \label{eq:Hnat}
H_{\rm int} = g \sum_{n=1}^{N_a} \sum_{j=1}^{p_n} \mleft[ \sigma_n^{(1), +} a_{x_{nj}} + \sigma_n^{(2), +} a_{x_{nj}} + \text{H.c.} \mright].
\eeq
Here, two-photon emission only becomes possible (through successive single-photon emission) when an atom is initialized in its $\ket{2}$ state, making the emergence of doublon BICs less intuitive and more conceptually rich.

To form a doublon BIC under the interaction in \eqref{eq:Hnat}, the energy of the $\ket{2}$ level must match the doublon dispersion, i.e., $\Delta_1 + \Delta_2 \sim U$ [see \figpanel{fig:setup}{b}]. Additionally, to suppress single-photon emission into the waveguide, both transition energies must lie outside the single-photon band: $|\Delta_{1,2}|>2J$. In this regime, the $\ket{2}$ state of the atom couples to doublon modes via a second-order process with an effective coupling strength $g_\text{eff}\sim g^2/\Delta_2$~\cite{footnote1} to the doublon modes in the waveguide. This effective coupling is essential to the formation of doublon BICs in this setup. An additional effect of the second-order process is that it adds a Lamb shift to the energy of the $\ket{2}$ state and to the doublon dispersion~\cite{footnote1}.

At a decoherence-free frequency $\omega_\text{DF} = \Delta_1 + \Delta_2 + \Delta$, this second-order process leads to the formation of a doublon BIC, as demonstrated in \figpanel{fig3}{a} for a single GA with parameters $g = 0.25 J$, $\Delta_1 = 5.338 J$, $\Delta_2 = 5 J$, and $\Delta = 0.054 J$; the $\Delta$ is due to the Lamb shift in the atomic levels~\cite{footnote1}. To confirm the BIC, we examine, in \figpanel{fig3}{b}, the $\ket{2}$-level occupation $n^{(2)}(t) = \bra{\psi(t)}\sigma^{(2), +}\sigma^{(2), -}\ket{\psi(t)}/2$ for different atomic frequencies, , starting from $\ket{\psi (t=0)} = \sigma^{(2),+}\sigma^{(1),+}\ket{\text{vac}}/\sqrt{2}$. At $\Delta_1+\Delta_2+\Delta=\omega_\text{DF}$ (blue curve), we observe a decoherence-free behavior where the population stabilizes at a high value. 
The dynamics for $\Delta_1+\Delta_2+\Delta\neq\omega_\text{DF}$ inside (orange curve) and outside (green curve) the band are similar to those in \figpanel{fig2}{b}, and distinct from the BIC behavior.


We demonstrate the occurrence of DFI under this single-photon coupling in \figpanels{fig3}{c}{d}, where, as in \figpanels{fig2}{c}{d}, we add a second GA coupled to cavities 100 and 102, forming a braided configuration. Starting from $\ket{\psi (t=0)} = \sigma_1^{(2),+}\sigma_1^{(1),+}\ket{\text{vac}}/\sqrt{2}$, we observe coherent doublon-mediated dynamics between two GAs in the same way as in the preceding section. We see that due to the weaker effective coupling, the exchange rate between the GAs is slower than in \figpanel{fig2}{c} (even though $g$ is larger in \figref{fig3} than in \figref{fig2}). Consequently, the doublon population being exchanged in \figpanel{fig3}{d} is smaller than that in \figpanel{fig2}{d}. Aside from these quantitative differences, the DFI behavior remains qualitatively the same as in the two-photon coupling case. A small detuning from the decoherence-free condition leads to the breakdown of DFI and leakage of doublons, as shown in~\figpanels{fig3}{e}{f}, further confirming the role of doublon BICs in mediating the DFI.





\textit{Conclusion and outlook.---}We have introduced giant atoms (GAs) coupled to non-linear structured waveguides as a platform to realize many-body bound states in the continuum (BICs), leveraging the destructive interference between spatially separated emission pathways---an effect unique to GAs. We first demonstrated the formation of doublon BICs in setups with direct two-photon emission to the waveguide and showed that these BICs can mediate decoherence-free interactions (DFIs) between different GAs. We then extended our analysis to the more experimentally relevant case where only single-photon emission is allowed and found that doublon BICs and their associated DFIs still emerge, enabled by virtual two-photon transitions. These findings highlight GAs as a versatile platform for realizing non-ergodic many-body states in open quantum systems and push forward interference as a powerful mechanism for engineering many-body BICs.

Experimentally, the required separation of doublon and free-photon dynamics (specifically, a nonlinear interaction strength $U\geq 4J$) is accessible in state-of-the-art platforms. In superconducting circuits, high-impedance resonators can yield $U/J\approx24$~\cite{Andersson2025}. Additionally, giant atoms have been realized with transmon qubits coupled to arrays of superconducting resonators~\cite{Jouanny2025}. In cold-atom systems, strong photon-photon interactions can be engineered via Rydberg blockade~\cite{PhysRevLett.107.133602,Peyronel2012}, and recent proposals show that GA configurations can be realized in these platforms as well~\cite{Gonzalez-Tudela2019}. The observation and control of doublon BICs is thus experimentally feasible across a range of architectures.

The non-radiative nature of doublon BICs makes them promising candidates for entangled-state preparation~\cite{Soro2022} and fault-tolerant quantum gates in noisy environments~\cite{Kannan2020, Chen2025, chen2025scalablequantumsimulatorextended, Chen2025b}. Their intrinsic many-body character enables applications beyond the single-excitation regime, such as preparing and distributing many-body entangled states~\cite{PhysRevLett.91.107903, PhysRevLett.81.5932, Wehner2018, RevModPhys.83.33}, and implementing unconventional quantum gates involving more than one photon~\cite{Caneva2015, Kimble2008}. Looking ahead, the principles demonstrated here naturally extend to $N\geq2$ many-body BICs~\cite{sugimoto2023manybodyboundstatescontinuum} in GAs with more levels. These states could facilitate protected multi-qubit entanglement, robust many-body state transfer, and novel localization phenomena in open quantum systems. The interplay between photon-photon interactions and interference thus opens a rich frontier for both theoretical and experimental exploration of many-body BICs. Finally, our strategy is not limited to bosonic systems such as the Bose--Hubbard model, but can also be extended to interacting fermionic systems. Examples include two-magnon bound states in Ising-like XXZ chains~\cite{PhysRevB.35.342, PhysRevB.105.064419, Takahashi1999}, two-fermion bound states in the one-dimensional Fermi--Hubbard model~\cite{Essler2005}, and more complex many-body excitations such as Cooper pairs.


\begin{acknowledgments}
    
\textit{Acknowledgments.---}All code used in this work is open access via GitHub: \hyperlink{https://github.com/WalterRieck/GiantAtoms}{https://github.com/WalterRieck/GiantAtoms}. We thank Ariadna Soro and Xin Wang for fruitful discussions. G.C.~is supported by the European Union's Horizon Europe programme HORIZON-MSCA-2023-PF-01-01 via the project 101146565 (SING-ATOM). A.F.K.~acknowledges support from the Swedish Foundation for Strategic Research (grant numbers FFL21-0279 and FUS21-0063), the Horizon Europe programme HORIZON-CL4-2022-QUANTUM-01-SGA via the project 101113946 OpenSuperQPlus100, and from the Knut and Alice Wallenberg Foundation through the Wallenberg Centre for Quantum Technology (WACQT). 

\end{acknowledgments}


%


\begin{thebibliography}{96}%
\makeatletter
\providecommand \@ifxundefined [1]{%
 \@ifx{#1\undefined}
}%
\providecommand \@ifnum [1]{%
 \ifnum #1\expandafter \@firstoftwo
 \else \expandafter \@secondoftwo
 \fi
}%
\providecommand \@ifx [1]{%
 \ifx #1\expandafter \@firstoftwo
 \else \expandafter \@secondoftwo
 \fi
}%
\providecommand \natexlab [1]{#1}%
\providecommand \enquote  [1]{``#1''}%
\providecommand \bibnamefont  [1]{#1}%
\providecommand \bibfnamefont [1]{#1}%
\providecommand \citenamefont [1]{#1}%
\providecommand \href@noop [0]{\@secondoftwo}%
\providecommand \href [0]{\begingroup \@sanitize@url \@href}%
\providecommand \@href[1]{\@@startlink{#1}\@@href}%
\providecommand \@@href[1]{\endgroup#1\@@endlink}%
\providecommand \@sanitize@url [0]{\catcode `\\12\catcode `\$12\catcode `\&12\catcode `\#12\catcode `\^12\catcode `\_12\catcode `\%12\relax}%
\providecommand \@@startlink[1]{}%
\providecommand \@@endlink[0]{}%
\providecommand \url  [0]{\begingroup\@sanitize@url \@url }%
\providecommand \@url [1]{\endgroup\@href {#1}{\urlprefix }}%
\providecommand \urlprefix  [0]{URL }%
\providecommand \Eprint [0]{\href }%
\providecommand \doibase [0]{https://doi.org/}%
\providecommand \selectlanguage [0]{\@gobble}%
\providecommand \bibinfo  [0]{\@secondoftwo}%
\providecommand \bibfield  [0]{\@secondoftwo}%
\providecommand \translation [1]{[#1]}%
\providecommand \BibitemOpen [0]{}%
\providecommand \bibitemStop [0]{}%
\providecommand \bibitemNoStop [0]{.\EOS\space}%
\providecommand \EOS [0]{\spacefactor3000\relax}%
\providecommand \BibitemShut  [1]{\csname bibitem#1\endcsname}%
\let\auto@bib@innerbib\@empty
\bibitem [{\citenamefont {von Neumann}\ and\ \citenamefont {Wigner}(1993)}]{vonNeumann1993}%
  \BibitemOpen
  \bibfield  {author} {\bibinfo {author} {\bibfnamefont {J.}~\bibnamefont {von Neumann}}\ and\ \bibinfo {author} {\bibfnamefont {E.~P.}\ \bibnamefont {Wigner}},\ }\bibinfo {title} {\"{U}ber merkw\"{u}rdige diskrete eigenwerte},\ in\ \href {https://doi.org/10.1007/978-3-662-02781-3_19} {\emph {\bibinfo {booktitle} {The Collected Works of Eugene Paul Wigner}}}\ (\bibinfo  {publisher} {Springer Berlin Heidelberg},\ \bibinfo {year} {1993})\ p.\ \bibinfo {pages} {291–293}\BibitemShut {NoStop}%
\bibitem [{\citenamefont {Stillinger}\ and\ \citenamefont {Herrick}(1975)}]{PhysRevA.11.446}%
  \BibitemOpen
  \bibfield  {author} {\bibinfo {author} {\bibfnamefont {F.~H.}\ \bibnamefont {Stillinger}}\ and\ \bibinfo {author} {\bibfnamefont {D.~R.}\ \bibnamefont {Herrick}},\ }\bibfield  {title} {\bibinfo {title} {Bound states in the continuum},\ }\href {https://doi.org/10.1103/PhysRevA.11.446} {\bibfield  {journal} {\bibinfo  {journal} {Physical Review A}\ }\textbf {\bibinfo {volume} {11}},\ \bibinfo {pages} {446} (\bibinfo {year} {1975})}\BibitemShut {NoStop}%
\bibitem [{\citenamefont {Plotnik}\ \emph {et~al.}(2011)\citenamefont {Plotnik}, \citenamefont {Peleg}, \citenamefont {Dreisow}, \citenamefont {Heinrich}, \citenamefont {Nolte}, \citenamefont {Szameit},\ and\ \citenamefont {Segev}}]{PhysRevLett.107.183901}%
  \BibitemOpen
  \bibfield  {author} {\bibinfo {author} {\bibfnamefont {Y.}~\bibnamefont {Plotnik}}, \bibinfo {author} {\bibfnamefont {O.}~\bibnamefont {Peleg}}, \bibinfo {author} {\bibfnamefont {F.}~\bibnamefont {Dreisow}}, \bibinfo {author} {\bibfnamefont {M.}~\bibnamefont {Heinrich}}, \bibinfo {author} {\bibfnamefont {S.}~\bibnamefont {Nolte}}, \bibinfo {author} {\bibfnamefont {A.}~\bibnamefont {Szameit}},\ and\ \bibinfo {author} {\bibfnamefont {M.}~\bibnamefont {Segev}},\ }\bibfield  {title} {\bibinfo {title} {{Experimental Observation of Optical Bound States in the Continuum}},\ }\href {https://doi.org/10.1103/PhysRevLett.107.183901} {\bibfield  {journal} {\bibinfo  {journal} {Physical Review Letters}\ }\textbf {\bibinfo {volume} {107}},\ \bibinfo {pages} {183901} (\bibinfo {year} {2011})}\BibitemShut {NoStop}%
\bibitem [{\citenamefont {Robnik}(1986)}]{Robnik1986}%
  \BibitemOpen
  \bibfield  {author} {\bibinfo {author} {\bibfnamefont {M.}~\bibnamefont {Robnik}},\ }\bibfield  {title} {\bibinfo {title} {{A simple separable Hamiltonian having bound states in the continuum}},\ }\href {https://doi.org/10.1088/0305-4470/19/18/029} {\bibfield  {journal} {\bibinfo  {journal} {Journal of Physics A: Mathematical and General}\ }\textbf {\bibinfo {volume} {19}},\ \bibinfo {pages} {3845} (\bibinfo {year} {1986})}\BibitemShut {NoStop}%
\bibitem [{\citenamefont {Friedrich}\ and\ \citenamefont {Wintgen}(1985)}]{PhysRevA.32.3231}%
  \BibitemOpen
  \bibfield  {author} {\bibinfo {author} {\bibfnamefont {H.}~\bibnamefont {Friedrich}}\ and\ \bibinfo {author} {\bibfnamefont {D.}~\bibnamefont {Wintgen}},\ }\bibfield  {title} {\bibinfo {title} {Interfering resonances and bound states in the continuum},\ }\href {https://doi.org/10.1103/PhysRevA.32.3231} {\bibfield  {journal} {\bibinfo  {journal} {Physical Review A}\ }\textbf {\bibinfo {volume} {32}},\ \bibinfo {pages} {3231} (\bibinfo {year} {1985})}\BibitemShut {NoStop}%
\bibitem [{\citenamefont {Marinica}\ \emph {et~al.}(2008)\citenamefont {Marinica}, \citenamefont {Borisov},\ and\ \citenamefont {Shabanov}}]{PhysRevLett.100.183902}%
  \BibitemOpen
  \bibfield  {author} {\bibinfo {author} {\bibfnamefont {D.~C.}\ \bibnamefont {Marinica}}, \bibinfo {author} {\bibfnamefont {A.~G.}\ \bibnamefont {Borisov}},\ and\ \bibinfo {author} {\bibfnamefont {S.~V.}\ \bibnamefont {Shabanov}},\ }\bibfield  {title} {\bibinfo {title} {{Bound States in the Continuum in Photonics}},\ }\href {https://doi.org/10.1103/PhysRevLett.100.183902} {\bibfield  {journal} {\bibinfo  {journal} {Physical Review Letters}\ }\textbf {\bibinfo {volume} {100}},\ \bibinfo {pages} {183902} (\bibinfo {year} {2008})}\BibitemShut {NoStop}%
\bibitem [{\citenamefont {Yang}\ \emph {et~al.}(2014)\citenamefont {Yang}, \citenamefont {Peng}, \citenamefont {Liang}, \citenamefont {Li},\ and\ \citenamefont {Noda}}]{PhysRevLett.113.037401}%
  \BibitemOpen
  \bibfield  {author} {\bibinfo {author} {\bibfnamefont {Y.}~\bibnamefont {Yang}}, \bibinfo {author} {\bibfnamefont {C.}~\bibnamefont {Peng}}, \bibinfo {author} {\bibfnamefont {Y.}~\bibnamefont {Liang}}, \bibinfo {author} {\bibfnamefont {Z.}~\bibnamefont {Li}},\ and\ \bibinfo {author} {\bibfnamefont {S.}~\bibnamefont {Noda}},\ }\bibfield  {title} {\bibinfo {title} {{Analytical Perspective for Bound States in the Continuum in Photonic Crystal Slabs}},\ }\href {https://doi.org/10.1103/PhysRevLett.113.037401} {\bibfield  {journal} {\bibinfo  {journal} {Physical Review Letters}\ }\textbf {\bibinfo {volume} {113}},\ \bibinfo {pages} {037401} (\bibinfo {year} {2014})}\BibitemShut {NoStop}%
\bibitem [{\citenamefont {Hsu}\ \emph {et~al.}(2016)\citenamefont {Hsu}, \citenamefont {Zhen}, \citenamefont {Stone}, \citenamefont {Joannopoulos},\ and\ \citenamefont {Soljačić}}]{Hsu2016}%
  \BibitemOpen
  \bibfield  {author} {\bibinfo {author} {\bibfnamefont {C.~W.}\ \bibnamefont {Hsu}}, \bibinfo {author} {\bibfnamefont {B.}~\bibnamefont {Zhen}}, \bibinfo {author} {\bibfnamefont {A.~D.}\ \bibnamefont {Stone}}, \bibinfo {author} {\bibfnamefont {J.~D.}\ \bibnamefont {Joannopoulos}},\ and\ \bibinfo {author} {\bibfnamefont {M.}~\bibnamefont {Soljačić}},\ }\bibfield  {title} {\bibinfo {title} {Bound states in the continuum},\ }\href {https://doi.org/10.1038/natrevmats.2016.48} {\bibfield  {journal} {\bibinfo  {journal} {Nature Reviews Materials}\ }\textbf {\bibinfo {volume} {1}},\ \bibinfo {pages} {16048} (\bibinfo {year} {2016})}\BibitemShut {NoStop}%
\bibitem [{\citenamefont {Koshelev}\ \emph {et~al.}(2020)\citenamefont {Koshelev}, \citenamefont {Bogdanov},\ and\ \citenamefont {Kivshar}}]{Koshelev2020}%
  \BibitemOpen
  \bibfield  {author} {\bibinfo {author} {\bibfnamefont {K.}~\bibnamefont {Koshelev}}, \bibinfo {author} {\bibfnamefont {A.}~\bibnamefont {Bogdanov}},\ and\ \bibinfo {author} {\bibfnamefont {Y.}~\bibnamefont {Kivshar}},\ }\bibfield  {title} {\bibinfo {title} {{Engineering with Bound States in the Continuum}},\ }\href {https://doi.org/10.1364/opn.31.1.000038} {\bibfield  {journal} {\bibinfo  {journal} {Optics and Photonics News}\ }\textbf {\bibinfo {volume} {31}},\ \bibinfo {pages} {38} (\bibinfo {year} {2020})}\BibitemShut {NoStop}%
\bibitem [{\citenamefont {Joseph}\ \emph {et~al.}(2021)\citenamefont {Joseph}, \citenamefont {Pandey}, \citenamefont {Sarkar},\ and\ \citenamefont {Joseph}}]{Joseph2021}%
  \BibitemOpen
  \bibfield  {author} {\bibinfo {author} {\bibfnamefont {S.}~\bibnamefont {Joseph}}, \bibinfo {author} {\bibfnamefont {S.}~\bibnamefont {Pandey}}, \bibinfo {author} {\bibfnamefont {S.}~\bibnamefont {Sarkar}},\ and\ \bibinfo {author} {\bibfnamefont {J.}~\bibnamefont {Joseph}},\ }\bibfield  {title} {\bibinfo {title} {Bound states in the continuum in resonant nanostructures: an overview of engineered materials for tailored applications},\ }\href {https://doi.org/10.1515/nanoph-2021-0387} {\bibfield  {journal} {\bibinfo  {journal} {Nanophotonics}\ }\textbf {\bibinfo {volume} {10}},\ \bibinfo {pages} {4175} (\bibinfo {year} {2021})}\BibitemShut {NoStop}%
\bibitem [{\citenamefont {Azzam}\ and\ \citenamefont {Kildishev}(2021)}]{Azzam2020}%
  \BibitemOpen
  \bibfield  {author} {\bibinfo {author} {\bibfnamefont {S.~I.}\ \bibnamefont {Azzam}}\ and\ \bibinfo {author} {\bibfnamefont {A.~V.}\ \bibnamefont {Kildishev}},\ }\bibfield  {title} {\bibinfo {title} {{Photonic Bound States in the Continuum: From Basics to Applications}},\ }\href {https://doi.org/10.1002/adom.202001469} {\bibfield  {journal} {\bibinfo  {journal} {Advanced Optical Materials}\ }\textbf {\bibinfo {volume} {9}},\ \bibinfo {pages} {2001469} (\bibinfo {year} {2021})}\BibitemShut {NoStop}%
\bibitem [{\citenamefont {Deriy}\ \emph {et~al.}(2022)\citenamefont {Deriy}, \citenamefont {Toftul}, \citenamefont {Petrov},\ and\ \citenamefont {Bogdanov}}]{PhysRevLett.128.084301}%
  \BibitemOpen
  \bibfield  {author} {\bibinfo {author} {\bibfnamefont {I.}~\bibnamefont {Deriy}}, \bibinfo {author} {\bibfnamefont {I.}~\bibnamefont {Toftul}}, \bibinfo {author} {\bibfnamefont {M.}~\bibnamefont {Petrov}},\ and\ \bibinfo {author} {\bibfnamefont {A.}~\bibnamefont {Bogdanov}},\ }\bibfield  {title} {\bibinfo {title} {{Bound States in the Continuum in Compact Acoustic Resonators}},\ }\href {https://doi.org/10.1103/PhysRevLett.128.084301} {\bibfield  {journal} {\bibinfo  {journal} {Physical Review Letters}\ }\textbf {\bibinfo {volume} {128}},\ \bibinfo {pages} {084301} (\bibinfo {year} {2022})}\BibitemShut {NoStop}%
\bibitem [{\citenamefont {Tong}\ \emph {et~al.}(2020)\citenamefont {Tong}, \citenamefont {Liu}, \citenamefont {Zhao},\ and\ \citenamefont {Fang}}]{Tong2020}%
  \BibitemOpen
  \bibfield  {author} {\bibinfo {author} {\bibfnamefont {H.}~\bibnamefont {Tong}}, \bibinfo {author} {\bibfnamefont {S.}~\bibnamefont {Liu}}, \bibinfo {author} {\bibfnamefont {M.}~\bibnamefont {Zhao}},\ and\ \bibinfo {author} {\bibfnamefont {K.}~\bibnamefont {Fang}},\ }\bibfield  {title} {\bibinfo {title} {Observation of phonon trapping in the continuum with topological charges},\ }\href {https://doi.org/10.1038/s41467-020-19091-3} {\bibfield  {journal} {\bibinfo  {journal} {Nature Communications}\ }\textbf {\bibinfo {volume} {11}},\ \bibinfo {pages} {5216} (\bibinfo {year} {2020})}\BibitemShut {NoStop}%
\bibitem [{\citenamefont {Carletti}\ \emph {et~al.}(2018)\citenamefont {Carletti}, \citenamefont {Koshelev}, \citenamefont {De~Angelis},\ and\ \citenamefont {Kivshar}}]{PhysRevLett.121.033903}%
  \BibitemOpen
  \bibfield  {author} {\bibinfo {author} {\bibfnamefont {L.}~\bibnamefont {Carletti}}, \bibinfo {author} {\bibfnamefont {K.}~\bibnamefont {Koshelev}}, \bibinfo {author} {\bibfnamefont {C.}~\bibnamefont {De~Angelis}},\ and\ \bibinfo {author} {\bibfnamefont {Y.}~\bibnamefont {Kivshar}},\ }\bibfield  {title} {\bibinfo {title} {{Giant Nonlinear Response at the Nanoscale Driven by Bound States in the Continuum}},\ }\href {https://doi.org/10.1103/PhysRevLett.121.033903} {\bibfield  {journal} {\bibinfo  {journal} {Physical Review Letters}\ }\textbf {\bibinfo {volume} {121}},\ \bibinfo {pages} {033903} (\bibinfo {year} {2018})}\BibitemShut {NoStop}%
\bibitem [{\citenamefont {Liu}\ \emph {et~al.}(2019)\citenamefont {Liu}, \citenamefont {Xu}, \citenamefont {Lin}, \citenamefont {Xiang}, \citenamefont {Feng}, \citenamefont {Cao}, \citenamefont {Li}, \citenamefont {Lan},\ and\ \citenamefont {Liu}}]{PhysRevLett.123.253901}%
  \BibitemOpen
  \bibfield  {author} {\bibinfo {author} {\bibfnamefont {Z.}~\bibnamefont {Liu}}, \bibinfo {author} {\bibfnamefont {Y.}~\bibnamefont {Xu}}, \bibinfo {author} {\bibfnamefont {Y.}~\bibnamefont {Lin}}, \bibinfo {author} {\bibfnamefont {J.}~\bibnamefont {Xiang}}, \bibinfo {author} {\bibfnamefont {T.}~\bibnamefont {Feng}}, \bibinfo {author} {\bibfnamefont {Q.}~\bibnamefont {Cao}}, \bibinfo {author} {\bibfnamefont {J.}~\bibnamefont {Li}}, \bibinfo {author} {\bibfnamefont {S.}~\bibnamefont {Lan}},\ and\ \bibinfo {author} {\bibfnamefont {J.}~\bibnamefont {Liu}},\ }\bibfield  {title} {\bibinfo {title} {{High-$Q$ Quasibound States in the Continuum for Nonlinear Metasurfaces}},\ }\href {https://doi.org/10.1103/PhysRevLett.123.253901} {\bibfield  {journal} {\bibinfo  {journal} {Physical Review Letters}\ }\textbf {\bibinfo {volume} {123}},\ \bibinfo {pages} {253901} (\bibinfo {year} {2019})}\BibitemShut {NoStop}%
\bibitem [{\citenamefont {Koshelev}\ \emph {et~al.}(2019)\citenamefont {Koshelev}, \citenamefont {Tang}, \citenamefont {Li}, \citenamefont {Choi}, \citenamefont {Li},\ and\ \citenamefont {Kivshar}}]{Koshelev2019}%
  \BibitemOpen
  \bibfield  {author} {\bibinfo {author} {\bibfnamefont {K.}~\bibnamefont {Koshelev}}, \bibinfo {author} {\bibfnamefont {Y.}~\bibnamefont {Tang}}, \bibinfo {author} {\bibfnamefont {K.}~\bibnamefont {Li}}, \bibinfo {author} {\bibfnamefont {D.-Y.}\ \bibnamefont {Choi}}, \bibinfo {author} {\bibfnamefont {G.}~\bibnamefont {Li}},\ and\ \bibinfo {author} {\bibfnamefont {Y.}~\bibnamefont {Kivshar}},\ }\bibfield  {title} {\bibinfo {title} {{Nonlinear Metasurfaces Governed by Bound States in the Continuum}},\ }\href {https://doi.org/10.1021/acsphotonics.9b00700} {\bibfield  {journal} {\bibinfo  {journal} {ACS Photonics}\ }\textbf {\bibinfo {volume} {6}},\ \bibinfo {pages} {1639} (\bibinfo {year} {2019})}\BibitemShut {NoStop}%
\bibitem [{\citenamefont {Carletti}\ \emph {et~al.}(2019)\citenamefont {Carletti}, \citenamefont {Kruk}, \citenamefont {Bogdanov}, \citenamefont {De~Angelis},\ and\ \citenamefont {Kivshar}}]{PhysRevResearch.1.023016}%
  \BibitemOpen
  \bibfield  {author} {\bibinfo {author} {\bibfnamefont {L.}~\bibnamefont {Carletti}}, \bibinfo {author} {\bibfnamefont {S.~S.}\ \bibnamefont {Kruk}}, \bibinfo {author} {\bibfnamefont {A.~A.}\ \bibnamefont {Bogdanov}}, \bibinfo {author} {\bibfnamefont {C.}~\bibnamefont {De~Angelis}},\ and\ \bibinfo {author} {\bibfnamefont {Y.}~\bibnamefont {Kivshar}},\ }\bibfield  {title} {\bibinfo {title} {High-harmonic generation at the nanoscale boosted by bound states in the continuum},\ }\href {https://doi.org/10.1103/PhysRevResearch.1.023016} {\bibfield  {journal} {\bibinfo  {journal} {Physical Review Research}\ }\textbf {\bibinfo {volume} {1}},\ \bibinfo {pages} {023016} (\bibinfo {year} {2019})}\BibitemShut {NoStop}%
\bibitem [{\citenamefont {Anthur}\ \emph {et~al.}(2020)\citenamefont {Anthur}, \citenamefont {Zhang}, \citenamefont {Paniagua-Dominguez}, \citenamefont {Kalashnikov}, \citenamefont {Ha}, \citenamefont {Maß}, \citenamefont {Kuznetsov},\ and\ \citenamefont {Krivitsky}}]{Anthur2020}%
  \BibitemOpen
  \bibfield  {author} {\bibinfo {author} {\bibfnamefont {A.~P.}\ \bibnamefont {Anthur}}, \bibinfo {author} {\bibfnamefont {H.}~\bibnamefont {Zhang}}, \bibinfo {author} {\bibfnamefont {R.}~\bibnamefont {Paniagua-Dominguez}}, \bibinfo {author} {\bibfnamefont {D.~A.}\ \bibnamefont {Kalashnikov}}, \bibinfo {author} {\bibfnamefont {S.~T.}\ \bibnamefont {Ha}}, \bibinfo {author} {\bibfnamefont {T.~W.~W.}\ \bibnamefont {Maß}}, \bibinfo {author} {\bibfnamefont {A.~I.}\ \bibnamefont {Kuznetsov}},\ and\ \bibinfo {author} {\bibfnamefont {L.}~\bibnamefont {Krivitsky}},\ }\bibfield  {title} {\bibinfo {title} {{Continuous Wave Second Harmonic Generation Enabled by Quasi-Bound-States in the Continuum on Gallium Phosphide Metasurfaces}},\ }\href {https://doi.org/10.1021/acs.nanolett.0c03601} {\bibfield  {journal} {\bibinfo  {journal} {Nano Letters}\ }\textbf {\bibinfo {volume} {20}},\ \bibinfo {pages} {8745} (\bibinfo {year} {2020})}\BibitemShut {NoStop}%
\bibitem [{\citenamefont {Zograf}\ \emph {et~al.}(2022)\citenamefont {Zograf}, \citenamefont {Koshelev}, \citenamefont {Zalogina}, \citenamefont {Korolev}, \citenamefont {Hollinger}, \citenamefont {Choi}, \citenamefont {Zuerch}, \citenamefont {Spielmann}, \citenamefont {Luther-Davies}, \citenamefont {Kartashov}, \citenamefont {Makarov}, \citenamefont {Kruk},\ and\ \citenamefont {Kivshar}}]{Zograf2022}%
  \BibitemOpen
  \bibfield  {author} {\bibinfo {author} {\bibfnamefont {G.}~\bibnamefont {Zograf}}, \bibinfo {author} {\bibfnamefont {K.}~\bibnamefont {Koshelev}}, \bibinfo {author} {\bibfnamefont {A.}~\bibnamefont {Zalogina}}, \bibinfo {author} {\bibfnamefont {V.}~\bibnamefont {Korolev}}, \bibinfo {author} {\bibfnamefont {R.}~\bibnamefont {Hollinger}}, \bibinfo {author} {\bibfnamefont {D.-Y.}\ \bibnamefont {Choi}}, \bibinfo {author} {\bibfnamefont {M.}~\bibnamefont {Zuerch}}, \bibinfo {author} {\bibfnamefont {C.}~\bibnamefont {Spielmann}}, \bibinfo {author} {\bibfnamefont {B.}~\bibnamefont {Luther-Davies}}, \bibinfo {author} {\bibfnamefont {D.}~\bibnamefont {Kartashov}}, \bibinfo {author} {\bibfnamefont {S.~V.}\ \bibnamefont {Makarov}}, \bibinfo {author} {\bibfnamefont {S.~S.}\ \bibnamefont {Kruk}},\ and\ \bibinfo {author} {\bibfnamefont {Y.}~\bibnamefont {Kivshar}},\ }\bibfield  {title} {\bibinfo {title} {{High-Harmonic Generation from Resonant Dielectric Metasurfaces Empowered by Bound States in the Continuum}},\ }\href
  {https://doi.org/10.1021/acsphotonics.1c01511} {\bibfield  {journal} {\bibinfo  {journal} {ACS Photonics}\ }\textbf {\bibinfo {volume} {9}},\ \bibinfo {pages} {567} (\bibinfo {year} {2022})}\BibitemShut {NoStop}%
\bibitem [{\citenamefont {Kodigala}\ \emph {et~al.}(2017)\citenamefont {Kodigala}, \citenamefont {Lepetit}, \citenamefont {Gu}, \citenamefont {Bahari}, \citenamefont {Fainman},\ and\ \citenamefont {Kanté}}]{Kodigala2017}%
  \BibitemOpen
  \bibfield  {author} {\bibinfo {author} {\bibfnamefont {A.}~\bibnamefont {Kodigala}}, \bibinfo {author} {\bibfnamefont {T.}~\bibnamefont {Lepetit}}, \bibinfo {author} {\bibfnamefont {Q.}~\bibnamefont {Gu}}, \bibinfo {author} {\bibfnamefont {B.}~\bibnamefont {Bahari}}, \bibinfo {author} {\bibfnamefont {Y.}~\bibnamefont {Fainman}},\ and\ \bibinfo {author} {\bibfnamefont {B.}~\bibnamefont {Kanté}},\ }\bibfield  {title} {\bibinfo {title} {Lasing action from photonic bound states in continuum},\ }\href {https://doi.org/10.1038/nature20799} {\bibfield  {journal} {\bibinfo  {journal} {Nature}\ }\textbf {\bibinfo {volume} {541}},\ \bibinfo {pages} {196} (\bibinfo {year} {2017})}\BibitemShut {NoStop}%
\bibitem [{\citenamefont {Ha}\ \emph {et~al.}(2018)\citenamefont {Ha}, \citenamefont {Fu}, \citenamefont {Emani}, \citenamefont {Pan}, \citenamefont {Bakker}, \citenamefont {Paniagua-Domínguez},\ and\ \citenamefont {Kuznetsov}}]{Ha2018}%
  \BibitemOpen
  \bibfield  {author} {\bibinfo {author} {\bibfnamefont {S.~T.}\ \bibnamefont {Ha}}, \bibinfo {author} {\bibfnamefont {Y.~H.}\ \bibnamefont {Fu}}, \bibinfo {author} {\bibfnamefont {N.~K.}\ \bibnamefont {Emani}}, \bibinfo {author} {\bibfnamefont {Z.}~\bibnamefont {Pan}}, \bibinfo {author} {\bibfnamefont {R.~M.}\ \bibnamefont {Bakker}}, \bibinfo {author} {\bibfnamefont {R.}~\bibnamefont {Paniagua-Domínguez}},\ and\ \bibinfo {author} {\bibfnamefont {A.~I.}\ \bibnamefont {Kuznetsov}},\ }\bibfield  {title} {\bibinfo {title} {Directional lasing in resonant semiconductor nanoantenna arrays},\ }\href {https://doi.org/10.1038/s41565-018-0245-5} {\bibfield  {journal} {\bibinfo  {journal} {Nature Nanotechnology}\ }\textbf {\bibinfo {volume} {13}},\ \bibinfo {pages} {1042} (\bibinfo {year} {2018})}\BibitemShut {NoStop}%
\bibitem [{\citenamefont {Wu}\ \emph {et~al.}(2020)\citenamefont {Wu}, \citenamefont {Ha}, \citenamefont {Shendre}, \citenamefont {Durmusoglu}, \citenamefont {Koh}, \citenamefont {Abujetas}, \citenamefont {Sánchez-Gil}, \citenamefont {Paniagua-Domínguez}, \citenamefont {Demir},\ and\ \citenamefont {Kuznetsov}}]{Wu2020}%
  \BibitemOpen
  \bibfield  {author} {\bibinfo {author} {\bibfnamefont {M.}~\bibnamefont {Wu}}, \bibinfo {author} {\bibfnamefont {S.~T.}\ \bibnamefont {Ha}}, \bibinfo {author} {\bibfnamefont {S.}~\bibnamefont {Shendre}}, \bibinfo {author} {\bibfnamefont {E.~G.}\ \bibnamefont {Durmusoglu}}, \bibinfo {author} {\bibfnamefont {W.-K.}\ \bibnamefont {Koh}}, \bibinfo {author} {\bibfnamefont {D.~R.}\ \bibnamefont {Abujetas}}, \bibinfo {author} {\bibfnamefont {J.~A.}\ \bibnamefont {Sánchez-Gil}}, \bibinfo {author} {\bibfnamefont {R.}~\bibnamefont {Paniagua-Domínguez}}, \bibinfo {author} {\bibfnamefont {H.~V.}\ \bibnamefont {Demir}},\ and\ \bibinfo {author} {\bibfnamefont {A.~I.}\ \bibnamefont {Kuznetsov}},\ }\bibfield  {title} {\bibinfo {title} {{Room-Temperature Lasing in Colloidal Nanoplatelets via Mie-Resonant Bound States in the Continuum}},\ }\href {https://doi.org/10.1021/acs.nanolett.0c01975} {\bibfield  {journal} {\bibinfo  {journal} {Nano Letters}\ }\textbf {\bibinfo {volume} {20}},\ \bibinfo {pages} {6005} (\bibinfo
  {year} {2020})}\BibitemShut {NoStop}%
\bibitem [{\citenamefont {Wang}\ \emph {et~al.}(2021{\natexlab{a}})\citenamefont {Wang}, \citenamefont {Fan}, \citenamefont {Zhang}, \citenamefont {Tang}, \citenamefont {Song}, \citenamefont {Han},\ and\ \citenamefont {Xiao}}]{Wang2021highly}%
  \BibitemOpen
  \bibfield  {author} {\bibinfo {author} {\bibfnamefont {Y.}~\bibnamefont {Wang}}, \bibinfo {author} {\bibfnamefont {Y.}~\bibnamefont {Fan}}, \bibinfo {author} {\bibfnamefont {X.}~\bibnamefont {Zhang}}, \bibinfo {author} {\bibfnamefont {H.}~\bibnamefont {Tang}}, \bibinfo {author} {\bibfnamefont {Q.}~\bibnamefont {Song}}, \bibinfo {author} {\bibfnamefont {J.}~\bibnamefont {Han}},\ and\ \bibinfo {author} {\bibfnamefont {S.}~\bibnamefont {Xiao}},\ }\bibfield  {title} {\bibinfo {title} {{Highly Controllable Etchless Perovskite Microlasers Based on Bound States in the Continuum}},\ }\href {https://doi.org/10.1021/acsnano.1c00673} {\bibfield  {journal} {\bibinfo  {journal} {ACS Nano}\ }\textbf {\bibinfo {volume} {15}},\ \bibinfo {pages} {7386} (\bibinfo {year} {2021}{\natexlab{a}})}\BibitemShut {NoStop}%
\bibitem [{\citenamefont {Yesilkoy}\ \emph {et~al.}(2019)\citenamefont {Yesilkoy}, \citenamefont {Arvelo}, \citenamefont {Jahani}, \citenamefont {Liu}, \citenamefont {Tittl}, \citenamefont {Cevher}, \citenamefont {Kivshar},\ and\ \citenamefont {Altug}}]{Yesilkoy2019}%
  \BibitemOpen
  \bibfield  {author} {\bibinfo {author} {\bibfnamefont {F.}~\bibnamefont {Yesilkoy}}, \bibinfo {author} {\bibfnamefont {E.~R.}\ \bibnamefont {Arvelo}}, \bibinfo {author} {\bibfnamefont {Y.}~\bibnamefont {Jahani}}, \bibinfo {author} {\bibfnamefont {M.}~\bibnamefont {Liu}}, \bibinfo {author} {\bibfnamefont {A.}~\bibnamefont {Tittl}}, \bibinfo {author} {\bibfnamefont {V.}~\bibnamefont {Cevher}}, \bibinfo {author} {\bibfnamefont {Y.}~\bibnamefont {Kivshar}},\ and\ \bibinfo {author} {\bibfnamefont {H.}~\bibnamefont {Altug}},\ }\bibfield  {title} {\bibinfo {title} {Ultrasensitive hyperspectral imaging and biodetection enabled by dielectric metasurfaces},\ }\href {https://doi.org/10.1038/s41566-019-0394-6} {\bibfield  {journal} {\bibinfo  {journal} {Nature Photonics}\ }\textbf {\bibinfo {volume} {13}},\ \bibinfo {pages} {390} (\bibinfo {year} {2019})}\BibitemShut {NoStop}%
\bibitem [{\citenamefont {Romano}\ \emph {et~al.}(2019)\citenamefont {Romano}, \citenamefont {Zito}, \citenamefont {Lara~Yépez}, \citenamefont {Cabrini}, \citenamefont {Penzo}, \citenamefont {Coppola}, \citenamefont {Rendina},\ and\ \citenamefont {Mocellaark}}]{Romano2019}%
  \BibitemOpen
  \bibfield  {author} {\bibinfo {author} {\bibfnamefont {S.}~\bibnamefont {Romano}}, \bibinfo {author} {\bibfnamefont {G.}~\bibnamefont {Zito}}, \bibinfo {author} {\bibfnamefont {S.~N.}\ \bibnamefont {Lara~Yépez}}, \bibinfo {author} {\bibfnamefont {S.}~\bibnamefont {Cabrini}}, \bibinfo {author} {\bibfnamefont {E.}~\bibnamefont {Penzo}}, \bibinfo {author} {\bibfnamefont {G.}~\bibnamefont {Coppola}}, \bibinfo {author} {\bibfnamefont {I.}~\bibnamefont {Rendina}},\ and\ \bibinfo {author} {\bibfnamefont {V.}~\bibnamefont {Mocellaark}},\ }\bibfield  {title} {\bibinfo {title} {Tuning the exponential sensitivity of a bound-state-in-continuum optical sensor},\ }\href {https://doi.org/10.1364/oe.27.018776} {\bibfield  {journal} {\bibinfo  {journal} {Optics Express}\ }\textbf {\bibinfo {volume} {27}},\ \bibinfo {pages} {18776} (\bibinfo {year} {2019})}\BibitemShut {NoStop}%
\bibitem [{\citenamefont {Tseng}\ \emph {et~al.}(2020)\citenamefont {Tseng}, \citenamefont {Jahani}, \citenamefont {Leitis},\ and\ \citenamefont {Altug}}]{Tseng2020}%
  \BibitemOpen
  \bibfield  {author} {\bibinfo {author} {\bibfnamefont {M.~L.}\ \bibnamefont {Tseng}}, \bibinfo {author} {\bibfnamefont {Y.}~\bibnamefont {Jahani}}, \bibinfo {author} {\bibfnamefont {A.}~\bibnamefont {Leitis}},\ and\ \bibinfo {author} {\bibfnamefont {H.}~\bibnamefont {Altug}},\ }\bibfield  {title} {\bibinfo {title} {{Dielectric Metasurfaces Enabling Advanced Optical Biosensors}},\ }\href {https://doi.org/10.1021/acsphotonics.0c01030} {\bibfield  {journal} {\bibinfo  {journal} {ACS Photonics}\ }\textbf {\bibinfo {volume} {8}},\ \bibinfo {pages} {47} (\bibinfo {year} {2020})}\BibitemShut {NoStop}%
\bibitem [{\citenamefont {Altug}\ \emph {et~al.}(2022)\citenamefont {Altug}, \citenamefont {Oh}, \citenamefont {Maier},\ and\ \citenamefont {Homola}}]{Altug2022}%
  \BibitemOpen
  \bibfield  {author} {\bibinfo {author} {\bibfnamefont {H.}~\bibnamefont {Altug}}, \bibinfo {author} {\bibfnamefont {S.-H.}\ \bibnamefont {Oh}}, \bibinfo {author} {\bibfnamefont {S.~A.}\ \bibnamefont {Maier}},\ and\ \bibinfo {author} {\bibfnamefont {J.}~\bibnamefont {Homola}},\ }\bibfield  {title} {\bibinfo {title} {Advances and applications of nanophotonic biosensors},\ }\href {https://doi.org/10.1038/s41565-021-01045-5} {\bibfield  {journal} {\bibinfo  {journal} {Nature Nanotechnology}\ }\textbf {\bibinfo {volume} {17}},\ \bibinfo {pages} {5} (\bibinfo {year} {2022})}\BibitemShut {NoStop}%
\bibitem [{\citenamefont {Chen}\ \emph {et~al.}(2020)\citenamefont {Chen}, \citenamefont {Zhao}, \citenamefont {Zhang},\ and\ \citenamefont {Qiu}}]{Chen2020}%
  \BibitemOpen
  \bibfield  {author} {\bibinfo {author} {\bibfnamefont {Y.}~\bibnamefont {Chen}}, \bibinfo {author} {\bibfnamefont {C.}~\bibnamefont {Zhao}}, \bibinfo {author} {\bibfnamefont {Y.}~\bibnamefont {Zhang}},\ and\ \bibinfo {author} {\bibfnamefont {C.-w.}\ \bibnamefont {Qiu}},\ }\bibfield  {title} {\bibinfo {title} {{Integrated Molar Chiral Sensing Based on High-Q Metasurface}},\ }\href {https://doi.org/10.1021/acs.nanolett.0c03506} {\bibfield  {journal} {\bibinfo  {journal} {Nano Letters}\ }\textbf {\bibinfo {volume} {20}},\ \bibinfo {pages} {8696} (\bibinfo {year} {2020})}\BibitemShut {NoStop}%
\bibitem [{\citenamefont {Sugimoto}\ \emph {et~al.}(2023)\citenamefont {Sugimoto}, \citenamefont {Ashida},\ and\ \citenamefont {Ueda}}]{sugimoto2023manybodyboundstatescontinuum}%
  \BibitemOpen
  \bibfield  {author} {\bibinfo {author} {\bibfnamefont {S.}~\bibnamefont {Sugimoto}}, \bibinfo {author} {\bibfnamefont {Y.}~\bibnamefont {Ashida}},\ and\ \bibinfo {author} {\bibfnamefont {M.}~\bibnamefont {Ueda}},\ }\href@noop {} {\bibinfo {title} {{Many-Body Bound States in the Continuum}}} (\bibinfo {year} {2023}),\ \Eprint {https://arxiv.org/abs/2307.05456} {arXiv:2307.05456} \BibitemShut {NoStop}%
\bibitem [{\citenamefont {Zhang}\ \emph {et~al.}(2012)\citenamefont {Zhang}, \citenamefont {Braak},\ and\ \citenamefont {Kollar}}]{PhysRevLett.109.116405}%
  \BibitemOpen
  \bibfield  {author} {\bibinfo {author} {\bibfnamefont {J.~M.}\ \bibnamefont {Zhang}}, \bibinfo {author} {\bibfnamefont {D.}~\bibnamefont {Braak}},\ and\ \bibinfo {author} {\bibfnamefont {M.}~\bibnamefont {Kollar}},\ }\bibfield  {title} {\bibinfo {title} {{Bound States in the Continuum Realized in the One-Dimensional Two-Particle Hubbard Model with an Impurity}},\ }\href {https://doi.org/10.1103/PhysRevLett.109.116405} {\bibfield  {journal} {\bibinfo  {journal} {Physical Review Letters}\ }\textbf {\bibinfo {volume} {109}},\ \bibinfo {pages} {116405} (\bibinfo {year} {2012})}\BibitemShut {NoStop}%
\bibitem [{\citenamefont {Ashida}\ \emph {et~al.}(2022)\citenamefont {Ashida}, \citenamefont {Yokota}, \citenamefont {\ifmmode \dot{I}\else \.{I}\fi{}mamo\ifmmode~\breve{g}\else \u{g}\fi{}lu},\ and\ \citenamefont {Demler}}]{PhysRevResearch.4.023194}%
  \BibitemOpen
  \bibfield  {author} {\bibinfo {author} {\bibfnamefont {Y.}~\bibnamefont {Ashida}}, \bibinfo {author} {\bibfnamefont {T.}~\bibnamefont {Yokota}}, \bibinfo {author} {\bibfnamefont {A.~m.~c.}\ \bibnamefont {\ifmmode \dot{I}\else \.{I}\fi{}mamo\ifmmode~\breve{g}\else \u{g}\fi{}lu}},\ and\ \bibinfo {author} {\bibfnamefont {E.}~\bibnamefont {Demler}},\ }\bibfield  {title} {\bibinfo {title} {Nonperturbative waveguide quantum electrodynamics},\ }\href {https://doi.org/10.1103/PhysRevResearch.4.023194} {\bibfield  {journal} {\bibinfo  {journal} {Physical Review Research}\ }\textbf {\bibinfo {volume} {4}},\ \bibinfo {pages} {023194} (\bibinfo {year} {2022})}\BibitemShut {NoStop}%
\bibitem [{\citenamefont {Sun}\ \emph {et~al.}(2024)\citenamefont {Sun}, \citenamefont {Zhang}, \citenamefont {Yuan},\ and\ \citenamefont {Zhang}}]{Sun2024}%
  \BibitemOpen
  \bibfield  {author} {\bibinfo {author} {\bibfnamefont {N.}~\bibnamefont {Sun}}, \bibinfo {author} {\bibfnamefont {W.}~\bibnamefont {Zhang}}, \bibinfo {author} {\bibfnamefont {H.}~\bibnamefont {Yuan}},\ and\ \bibinfo {author} {\bibfnamefont {X.}~\bibnamefont {Zhang}},\ }\bibfield  {title} {\bibinfo {title} {Boundary-localized many-body bound states in the continuum},\ }\href {https://doi.org/10.1038/s42005-024-01777-5} {\bibfield  {journal} {\bibinfo  {journal} {Communications Physics}\ }\textbf {\bibinfo {volume} {7}},\ \bibinfo {pages} {299} (\bibinfo {year} {2024})}\BibitemShut {NoStop}%
\bibitem [{\citenamefont {Kockum}(2021)}]{Kockum2021}%
  \BibitemOpen
  \bibfield  {author} {\bibinfo {author} {\bibfnamefont {A.~F.}\ \bibnamefont {Kockum}},\ }\bibfield  {title} {\bibinfo {title} {{Quantum Optics with Giant Atoms - the First Five Years}},\ }in\ \href {https://doi.org/10.1007/978-981-15-5191-8_12} {\emph {\bibinfo {booktitle} {International Symposium on Mathematics, Quantum Theory, and Cryptography (Mathematics for Industry, vol 33)}}}\ (\bibinfo  {publisher} {Springer},\ \bibinfo {year} {2021})\ pp.\ \bibinfo {pages} {125--146}\BibitemShut {NoStop}%
\bibitem [{\citenamefont {Frisk~Kockum}\ \emph {et~al.}(2014)\citenamefont {Frisk~Kockum}, \citenamefont {Delsing},\ and\ \citenamefont {Johansson}}]{Kockum2014}%
  \BibitemOpen
  \bibfield  {author} {\bibinfo {author} {\bibfnamefont {A.}~\bibnamefont {Frisk~Kockum}}, \bibinfo {author} {\bibfnamefont {P.}~\bibnamefont {Delsing}},\ and\ \bibinfo {author} {\bibfnamefont {G.}~\bibnamefont {Johansson}},\ }\bibfield  {title} {\bibinfo {title} {{Designing frequency-dependent relaxation rates and Lamb shifts for a giant artificial atom}},\ }\href {https://doi.org/10.1103/PhysRevA.90.013837} {\bibfield  {journal} {\bibinfo  {journal} {Physical Review A}\ }\textbf {\bibinfo {volume} {90}},\ \bibinfo {pages} {013837} (\bibinfo {year} {2014})}\BibitemShut {NoStop}%
\bibitem [{\citenamefont {Guo}\ \emph {et~al.}(2017)\citenamefont {Guo}, \citenamefont {Grimsmo}, \citenamefont {Kockum}, \citenamefont {Pletyukhov},\ and\ \citenamefont {Johansson}}]{Guo2017}%
  \BibitemOpen
  \bibfield  {author} {\bibinfo {author} {\bibfnamefont {L.}~\bibnamefont {Guo}}, \bibinfo {author} {\bibfnamefont {A.~L.}\ \bibnamefont {Grimsmo}}, \bibinfo {author} {\bibfnamefont {A.~F.}\ \bibnamefont {Kockum}}, \bibinfo {author} {\bibfnamefont {M.}~\bibnamefont {Pletyukhov}},\ and\ \bibinfo {author} {\bibfnamefont {G.}~\bibnamefont {Johansson}},\ }\bibfield  {title} {\bibinfo {title} {{Giant acoustic atom: A single quantum system with a deterministic time delay}},\ }\href {https://doi.org/10.1103/PhysRevA.95.053821} {\bibfield  {journal} {\bibinfo  {journal} {Physical Review A}\ }\textbf {\bibinfo {volume} {95}},\ \bibinfo {pages} {053821} (\bibinfo {year} {2017})}\BibitemShut {NoStop}%
\bibitem [{\citenamefont {Kockum}\ \emph {et~al.}(2018)\citenamefont {Kockum}, \citenamefont {Johansson},\ and\ \citenamefont {Nori}}]{Kockum2018}%
  \BibitemOpen
  \bibfield  {author} {\bibinfo {author} {\bibfnamefont {A.~F.}\ \bibnamefont {Kockum}}, \bibinfo {author} {\bibfnamefont {G.}~\bibnamefont {Johansson}},\ and\ \bibinfo {author} {\bibfnamefont {F.}~\bibnamefont {Nori}},\ }\bibfield  {title} {\bibinfo {title} {{Decoherence-Free Interaction between Giant Atoms in Waveguide Quantum Electrodynamics}},\ }\href {https://doi.org/10.1103/PhysRevLett.120.140404} {\bibfield  {journal} {\bibinfo  {journal} {Physical Review Letters}\ }\textbf {\bibinfo {volume} {120}},\ \bibinfo {pages} {140404} (\bibinfo {year} {2018})}\BibitemShut {NoStop}%
\bibitem [{\citenamefont {Gonz{\'{a}}lez-Tudela}\ \emph {et~al.}(2019)\citenamefont {Gonz{\'{a}}lez-Tudela}, \citenamefont {Mu{\~{n}}oz},\ and\ \citenamefont {Cirac}}]{Gonzalez-Tudela2019}%
  \BibitemOpen
  \bibfield  {author} {\bibinfo {author} {\bibfnamefont {A.}~\bibnamefont {Gonz{\'{a}}lez-Tudela}}, \bibinfo {author} {\bibfnamefont {C.~S.}\ \bibnamefont {Mu{\~{n}}oz}},\ and\ \bibinfo {author} {\bibfnamefont {J.~I.}\ \bibnamefont {Cirac}},\ }\bibfield  {title} {\bibinfo {title} {{Engineering and Harnessing Giant Atoms in High-Dimensional Baths: A Proposal for Implementation with Cold Atoms}},\ }\href {https://doi.org/10.1103/PhysRevLett.122.203603} {\bibfield  {journal} {\bibinfo  {journal} {Physical Review Letters}\ }\textbf {\bibinfo {volume} {122}},\ \bibinfo {pages} {203603} (\bibinfo {year} {2019})}\BibitemShut {NoStop}%
\bibitem [{\citenamefont {Guo}\ \emph {et~al.}(2020{\natexlab{a}})\citenamefont {Guo}, \citenamefont {Kockum}, \citenamefont {Marquardt},\ and\ \citenamefont {Johansson}}]{Guo2020}%
  \BibitemOpen
  \bibfield  {author} {\bibinfo {author} {\bibfnamefont {L.}~\bibnamefont {Guo}}, \bibinfo {author} {\bibfnamefont {A.~F.}\ \bibnamefont {Kockum}}, \bibinfo {author} {\bibfnamefont {F.}~\bibnamefont {Marquardt}},\ and\ \bibinfo {author} {\bibfnamefont {G.}~\bibnamefont {Johansson}},\ }\bibfield  {title} {\bibinfo {title} {{Oscillating bound states for a giant atom}},\ }\href {https://doi.org/10.1103/PhysRevResearch.2.043014} {\bibfield  {journal} {\bibinfo  {journal} {Physical Review Research}\ }\textbf {\bibinfo {volume} {2}},\ \bibinfo {pages} {043014} (\bibinfo {year} {2020}{\natexlab{a}})}\BibitemShut {NoStop}%
\bibitem [{\citenamefont {Guimond}\ \emph {et~al.}(2020)\citenamefont {Guimond}, \citenamefont {Vermersch}, \citenamefont {Juan}, \citenamefont {Sharafiev}, \citenamefont {Kirchmair},\ and\ \citenamefont {Zoller}}]{Guimond2020}%
  \BibitemOpen
  \bibfield  {author} {\bibinfo {author} {\bibfnamefont {P.~O.}\ \bibnamefont {Guimond}}, \bibinfo {author} {\bibfnamefont {B.}~\bibnamefont {Vermersch}}, \bibinfo {author} {\bibfnamefont {M.~L.}\ \bibnamefont {Juan}}, \bibinfo {author} {\bibfnamefont {A.}~\bibnamefont {Sharafiev}}, \bibinfo {author} {\bibfnamefont {G.}~\bibnamefont {Kirchmair}},\ and\ \bibinfo {author} {\bibfnamefont {P.}~\bibnamefont {Zoller}},\ }\bibfield  {title} {\bibinfo {title} {{A unidirectional on-chip photonic interface for superconducting circuits}},\ }\href {https://doi.org/10.1038/s41534-020-0261-9} {\bibfield  {journal} {\bibinfo  {journal} {npj Quantum Information}\ }\textbf {\bibinfo {volume} {6}},\ \bibinfo {pages} {32} (\bibinfo {year} {2020})}\BibitemShut {NoStop}%
\bibitem [{\citenamefont {Ask}\ \emph {et~al.}(2020)\citenamefont {Ask}, \citenamefont {Fang},\ and\ \citenamefont {Kockum}}]{Ask2020}%
  \BibitemOpen
  \bibfield  {author} {\bibinfo {author} {\bibfnamefont {A.}~\bibnamefont {Ask}}, \bibinfo {author} {\bibfnamefont {Y.-L.~L.}\ \bibnamefont {Fang}},\ and\ \bibinfo {author} {\bibfnamefont {A.~F.}\ \bibnamefont {Kockum}},\ }\href@noop {} {\bibinfo {title} {{Synthesizing electromagnetically induced transparency without a control field in waveguide QED using small and giant atoms}}} (\bibinfo {year} {2020}),\ \Eprint {https://arxiv.org/abs/2011.15077} {arXiv:2011.15077} \BibitemShut {NoStop}%
\bibitem [{\citenamefont {Cilluffo}\ \emph {et~al.}(2020)\citenamefont {Cilluffo}, \citenamefont {Carollo}, \citenamefont {Lorenzo}, \citenamefont {Gross}, \citenamefont {Palma},\ and\ \citenamefont {Ciccarello}}]{Cilluffo2020}%
  \BibitemOpen
  \bibfield  {author} {\bibinfo {author} {\bibfnamefont {D.}~\bibnamefont {Cilluffo}}, \bibinfo {author} {\bibfnamefont {A.}~\bibnamefont {Carollo}}, \bibinfo {author} {\bibfnamefont {S.}~\bibnamefont {Lorenzo}}, \bibinfo {author} {\bibfnamefont {J.~A.}\ \bibnamefont {Gross}}, \bibinfo {author} {\bibfnamefont {G.~M.}\ \bibnamefont {Palma}},\ and\ \bibinfo {author} {\bibfnamefont {F.}~\bibnamefont {Ciccarello}},\ }\bibfield  {title} {\bibinfo {title} {{Collisional picture of quantum optics with giant emitters}},\ }\href {https://doi.org/10.1103/PhysRevResearch.2.043070} {\bibfield  {journal} {\bibinfo  {journal} {Physical Review Research}\ }\textbf {\bibinfo {volume} {2}},\ \bibinfo {pages} {043070} (\bibinfo {year} {2020})}\BibitemShut {NoStop}%
\bibitem [{\citenamefont {Wang}\ \emph {et~al.}(2021{\natexlab{b}})\citenamefont {Wang}, \citenamefont {Liu}, \citenamefont {Kockum}, \citenamefont {Li},\ and\ \citenamefont {Nori}}]{Wang2021}%
  \BibitemOpen
  \bibfield  {author} {\bibinfo {author} {\bibfnamefont {X.}~\bibnamefont {Wang}}, \bibinfo {author} {\bibfnamefont {T.}~\bibnamefont {Liu}}, \bibinfo {author} {\bibfnamefont {A.~F.}\ \bibnamefont {Kockum}}, \bibinfo {author} {\bibfnamefont {H.-R.}\ \bibnamefont {Li}},\ and\ \bibinfo {author} {\bibfnamefont {F.}~\bibnamefont {Nori}},\ }\bibfield  {title} {\bibinfo {title} {{Tunable Chiral Bound States with Giant Atoms}},\ }\href {https://doi.org/10.1103/PhysRevLett.126.043602} {\bibfield  {journal} {\bibinfo  {journal} {Physical Review Letters}\ }\textbf {\bibinfo {volume} {126}},\ \bibinfo {pages} {043602} (\bibinfo {year} {2021}{\natexlab{b}})}\BibitemShut {NoStop}%
\bibitem [{\citenamefont {Du}\ and\ \citenamefont {Li}(2021)}]{Du2021}%
  \BibitemOpen
  \bibfield  {author} {\bibinfo {author} {\bibfnamefont {L.}~\bibnamefont {Du}}\ and\ \bibinfo {author} {\bibfnamefont {Y.}~\bibnamefont {Li}},\ }\bibfield  {title} {\bibinfo {title} {{Single-photon frequency conversion via a giant $\Lambda$-type atom}},\ }\href {https://doi.org/10.1103/PhysRevA.104.023712} {\bibfield  {journal} {\bibinfo  {journal} {Physical Review A}\ }\textbf {\bibinfo {volume} {104}},\ \bibinfo {pages} {023712} (\bibinfo {year} {2021})}\BibitemShut {NoStop}%
\bibitem [{\citenamefont {Soro}\ and\ \citenamefont {Kockum}(2022)}]{Soro2022}%
  \BibitemOpen
  \bibfield  {author} {\bibinfo {author} {\bibfnamefont {A.}~\bibnamefont {Soro}}\ and\ \bibinfo {author} {\bibfnamefont {A.~F.}\ \bibnamefont {Kockum}},\ }\bibfield  {title} {\bibinfo {title} {{Chiral quantum optics with giant atoms}},\ }\href {https://doi.org/10.1103/PhysRevA.105.023712} {\bibfield  {journal} {\bibinfo  {journal} {Physical Review A}\ }\textbf {\bibinfo {volume} {105}},\ \bibinfo {pages} {023712} (\bibinfo {year} {2022})}\BibitemShut {NoStop}%
\bibitem [{\citenamefont {Wang}\ and\ \citenamefont {Li}(2022)}]{Wang2022}%
  \BibitemOpen
  \bibfield  {author} {\bibinfo {author} {\bibfnamefont {X.}~\bibnamefont {Wang}}\ and\ \bibinfo {author} {\bibfnamefont {H.-r.}\ \bibnamefont {Li}},\ }\bibfield  {title} {\bibinfo {title} {{Chiral quantum network with giant atoms}},\ }\href {https://doi.org/10.1088/2058-9565/ac6a04} {\bibfield  {journal} {\bibinfo  {journal} {Quantum Science and Technology}\ }\textbf {\bibinfo {volume} {7}},\ \bibinfo {pages} {035007} (\bibinfo {year} {2022})}\BibitemShut {NoStop}%
\bibitem [{\citenamefont {Du}\ \emph {et~al.}(2022{\natexlab{a}})\citenamefont {Du}, \citenamefont {Zhang}, \citenamefont {Wu}, \citenamefont {Kockum},\ and\ \citenamefont {Li}}]{Du2022}%
  \BibitemOpen
  \bibfield  {author} {\bibinfo {author} {\bibfnamefont {L.}~\bibnamefont {Du}}, \bibinfo {author} {\bibfnamefont {Y.}~\bibnamefont {Zhang}}, \bibinfo {author} {\bibfnamefont {J.~H.}\ \bibnamefont {Wu}}, \bibinfo {author} {\bibfnamefont {A.~F.}\ \bibnamefont {Kockum}},\ and\ \bibinfo {author} {\bibfnamefont {Y.}~\bibnamefont {Li}},\ }\bibfield  {title} {\bibinfo {title} {{Giant Atoms in a Synthetic Frequency Dimension}},\ }\href {https://doi.org/10.1103/PhysRevLett.128.223602} {\bibfield  {journal} {\bibinfo  {journal} {Physical Review Letters}\ }\textbf {\bibinfo {volume} {128}},\ \bibinfo {pages} {223602} (\bibinfo {year} {2022}{\natexlab{a}})}\BibitemShut {NoStop}%
\bibitem [{\citenamefont {Du}\ \emph {et~al.}(2022{\natexlab{b}})\citenamefont {Du}, \citenamefont {Chen}, \citenamefont {Zhang},\ and\ \citenamefont {Li}}]{Du2022a}%
  \BibitemOpen
  \bibfield  {author} {\bibinfo {author} {\bibfnamefont {L.}~\bibnamefont {Du}}, \bibinfo {author} {\bibfnamefont {Y.-T.}\ \bibnamefont {Chen}}, \bibinfo {author} {\bibfnamefont {Y.}~\bibnamefont {Zhang}},\ and\ \bibinfo {author} {\bibfnamefont {Y.}~\bibnamefont {Li}},\ }\bibfield  {title} {\bibinfo {title} {{Giant atoms with time-dependent couplings}},\ }\href {https://doi.org/10.1103/PhysRevResearch.4.023198} {\bibfield  {journal} {\bibinfo  {journal} {Physical Review Research}\ }\textbf {\bibinfo {volume} {4}},\ \bibinfo {pages} {023198} (\bibinfo {year} {2022}{\natexlab{b}})}\BibitemShut {NoStop}%
\bibitem [{\citenamefont {Terradas-Brians{\'{o}}}\ \emph {et~al.}(2022)\citenamefont {Terradas-Brians{\'{o}}}, \citenamefont {Gonz{\'{a}}lez-Guti{\'{e}}rrez}, \citenamefont {Nori}, \citenamefont {Mart{\'{i}}n-Moreno},\ and\ \citenamefont {Zueco}}]{Terradas-Brianso2022}%
  \BibitemOpen
  \bibfield  {author} {\bibinfo {author} {\bibfnamefont {S.}~\bibnamefont {Terradas-Brians{\'{o}}}}, \bibinfo {author} {\bibfnamefont {C.~A.}\ \bibnamefont {Gonz{\'{a}}lez-Guti{\'{e}}rrez}}, \bibinfo {author} {\bibfnamefont {F.}~\bibnamefont {Nori}}, \bibinfo {author} {\bibfnamefont {L.}~\bibnamefont {Mart{\'{i}}n-Moreno}},\ and\ \bibinfo {author} {\bibfnamefont {D.}~\bibnamefont {Zueco}},\ }\bibfield  {title} {\bibinfo {title} {{Ultrastrong waveguide QED with giant atoms}},\ }\href {https://doi.org/10.1103/PhysRevA.106.063717} {\bibfield  {journal} {\bibinfo  {journal} {Physical Review A}\ }\textbf {\bibinfo {volume} {106}},\ \bibinfo {pages} {063717} (\bibinfo {year} {2022})}\BibitemShut {NoStop}%
\bibitem [{\citenamefont {Soro}\ \emph {et~al.}(2023)\citenamefont {Soro}, \citenamefont {Mu{\~{n}}oz},\ and\ \citenamefont {Kockum}}]{Soro2023}%
  \BibitemOpen
  \bibfield  {author} {\bibinfo {author} {\bibfnamefont {A.}~\bibnamefont {Soro}}, \bibinfo {author} {\bibfnamefont {C.~S.}\ \bibnamefont {Mu{\~{n}}oz}},\ and\ \bibinfo {author} {\bibfnamefont {A.~F.}\ \bibnamefont {Kockum}},\ }\bibfield  {title} {\bibinfo {title} {{Interaction between giant atoms in a one-dimensional structured environment}},\ }\href {https://doi.org/10.1103/PhysRevA.107.013710} {\bibfield  {journal} {\bibinfo  {journal} {Physical Review A}\ }\textbf {\bibinfo {volume} {107}},\ \bibinfo {pages} {013710} (\bibinfo {year} {2023})}\BibitemShut {NoStop}%
\bibitem [{\citenamefont {Du}\ \emph {et~al.}(2023)\citenamefont {Du}, \citenamefont {Guo}, \citenamefont {Zhang},\ and\ \citenamefont {Kockum}}]{Du2023}%
  \BibitemOpen
  \bibfield  {author} {\bibinfo {author} {\bibfnamefont {L.}~\bibnamefont {Du}}, \bibinfo {author} {\bibfnamefont {L.}~\bibnamefont {Guo}}, \bibinfo {author} {\bibfnamefont {Y.}~\bibnamefont {Zhang}},\ and\ \bibinfo {author} {\bibfnamefont {A.~F.}\ \bibnamefont {Kockum}},\ }\bibfield  {title} {\bibinfo {title} {{Giant emitters in a structured bath with non-Hermitian skin effect}},\ }\href {https://doi.org/10.1103/PhysRevResearch.5.L042040} {\bibfield  {journal} {\bibinfo  {journal} {Physical Review Research}\ }\textbf {\bibinfo {volume} {5}},\ \bibinfo {pages} {L042040} (\bibinfo {year} {2023})}\BibitemShut {NoStop}%
\bibitem [{\citenamefont {Ingelsten}\ \emph {et~al.}(2024)\citenamefont {Ingelsten}, \citenamefont {Kockum},\ and\ \citenamefont {Soro}}]{Ingelsten2024}%
  \BibitemOpen
  \bibfield  {author} {\bibinfo {author} {\bibfnamefont {E.~R.}\ \bibnamefont {Ingelsten}}, \bibinfo {author} {\bibfnamefont {A.~F.}\ \bibnamefont {Kockum}},\ and\ \bibinfo {author} {\bibfnamefont {A.}~\bibnamefont {Soro}},\ }\href@noop {} {\bibinfo {title} {{Avoiding decoherence with giant atoms in a two-dimensional structured environment}}} (\bibinfo {year} {2024}),\ \Eprint {https://arxiv.org/abs/2402.10879} {arXiv:2402.10879} \BibitemShut {NoStop}%
\bibitem [{\citenamefont {Leonforte}\ \emph {et~al.}(2025)\citenamefont {Leonforte}, \citenamefont {Sun}, \citenamefont {Valenti}, \citenamefont {Spagnolo}, \citenamefont {Illuminati}, \citenamefont {Carollo},\ and\ \citenamefont {Ciccarello}}]{Leonforte2024}%
  \BibitemOpen
  \bibfield  {author} {\bibinfo {author} {\bibfnamefont {L.}~\bibnamefont {Leonforte}}, \bibinfo {author} {\bibfnamefont {X.}~\bibnamefont {Sun}}, \bibinfo {author} {\bibfnamefont {D.}~\bibnamefont {Valenti}}, \bibinfo {author} {\bibfnamefont {B.}~\bibnamefont {Spagnolo}}, \bibinfo {author} {\bibfnamefont {F.}~\bibnamefont {Illuminati}}, \bibinfo {author} {\bibfnamefont {A.}~\bibnamefont {Carollo}},\ and\ \bibinfo {author} {\bibfnamefont {F.}~\bibnamefont {Ciccarello}},\ }\bibfield  {title} {\bibinfo {title} {{Quantum optics with giant atoms in a structured photonic bath}},\ }\href {https://doi.org/10.1088/2058-9565/ada08d} {\bibfield  {journal} {\bibinfo  {journal} {Quantum Science and Technology}\ }\textbf {\bibinfo {volume} {10}},\ \bibinfo {pages} {015057} (\bibinfo {year} {2025})}\BibitemShut {NoStop}%
\bibitem [{\citenamefont {Wang}\ \emph {et~al.}(2024{\natexlab{a}})\citenamefont {Wang}, \citenamefont {Zhu}, \citenamefont {Liu},\ and\ \citenamefont {Nori}}]{Wang2024}%
  \BibitemOpen
  \bibfield  {author} {\bibinfo {author} {\bibfnamefont {X.}~\bibnamefont {Wang}}, \bibinfo {author} {\bibfnamefont {H.-B.}\ \bibnamefont {Zhu}}, \bibinfo {author} {\bibfnamefont {T.}~\bibnamefont {Liu}},\ and\ \bibinfo {author} {\bibfnamefont {F.}~\bibnamefont {Nori}},\ }\bibfield  {title} {\bibinfo {title} {Realizing quantum optics in structured environments with giant atoms},\ }\href {https://doi.org/10.1103/PhysRevResearch.6.013279} {\bibfield  {journal} {\bibinfo  {journal} {Physical Review Research}\ }\textbf {\bibinfo {volume} {6}},\ \bibinfo {pages} {013279} (\bibinfo {year} {2024}{\natexlab{a}})}\BibitemShut {NoStop}%
\bibitem [{\citenamefont {Roccati}\ and\ \citenamefont {Cilluffo}(2024)}]{Roccati2024}%
  \BibitemOpen
  \bibfield  {author} {\bibinfo {author} {\bibfnamefont {F.}~\bibnamefont {Roccati}}\ and\ \bibinfo {author} {\bibfnamefont {D.}~\bibnamefont {Cilluffo}},\ }\bibfield  {title} {\bibinfo {title} {{Controlling Markovianity with Chiral Giant Atoms}},\ }\href {https://doi.org/10.1103/PhysRevLett.133.063603} {\bibfield  {journal} {\bibinfo  {journal} {Physical Review Letters}\ }\textbf {\bibinfo {volume} {133}},\ \bibinfo {pages} {063603} (\bibinfo {year} {2024})}\BibitemShut {NoStop}%
\bibitem [{\citenamefont {Gong}\ \emph {et~al.}(2024)\citenamefont {Gong}, \citenamefont {He}, \citenamefont {Yu}, \citenamefont {Zhang}, \citenamefont {Nori},\ and\ \citenamefont {Xiang}}]{Gong2024}%
  \BibitemOpen
  \bibfield  {author} {\bibinfo {author} {\bibfnamefont {R.-Y.}\ \bibnamefont {Gong}}, \bibinfo {author} {\bibfnamefont {Z.-Y.}\ \bibnamefont {He}}, \bibinfo {author} {\bibfnamefont {C.-H.}\ \bibnamefont {Yu}}, \bibinfo {author} {\bibfnamefont {G.-F.}\ \bibnamefont {Zhang}}, \bibinfo {author} {\bibfnamefont {F.}~\bibnamefont {Nori}},\ and\ \bibinfo {author} {\bibfnamefont {Z.-L.}\ \bibnamefont {Xiang}},\ }\href@noop {} {\bibinfo {title} {{Tunable quantum router with giant atoms, implementing quantum gates, teleportation, non-reciprocity, and circulators}}} (\bibinfo {year} {2024}),\ \Eprint {https://arxiv.org/abs/2411.19307} {arXiv:2411.19307} \BibitemShut {NoStop}%
\bibitem [{\citenamefont {Du}\ and\ \citenamefont {Kockum}(2025)}]{Du2025}%
  \BibitemOpen
  \bibfield  {author} {\bibinfo {author} {\bibfnamefont {L.}~\bibnamefont {Du}}\ and\ \bibinfo {author} {\bibfnamefont {A.~F.}\ \bibnamefont {Kockum}},\ }\bibfield  {title} {\bibinfo {title} {{Unconventional and robust light-matter interactions based on the non-Hermitian skin effect}},\ }\href {https://doi.org/10.1103/PhysRevResearch.7.013140} {\bibfield  {journal} {\bibinfo  {journal} {Physical Review Research}\ }\textbf {\bibinfo {volume} {7}},\ \bibinfo {pages} {013140} (\bibinfo {year} {2025})}\BibitemShut {NoStop}%
\bibitem [{\citenamefont {Du}\ \emph {et~al.}(2025)\citenamefont {Du}, \citenamefont {Wang}, \citenamefont {Kockum},\ and\ \citenamefont {Splettstoesser}}]{Du2025a}%
  \BibitemOpen
  \bibfield  {author} {\bibinfo {author} {\bibfnamefont {L.}~\bibnamefont {Du}}, \bibinfo {author} {\bibfnamefont {X.}~\bibnamefont {Wang}}, \bibinfo {author} {\bibfnamefont {A.~F.}\ \bibnamefont {Kockum}},\ and\ \bibinfo {author} {\bibfnamefont {J.}~\bibnamefont {Splettstoesser}},\ }\href@noop {} {\bibinfo {title} {{Dressed Interference in Giant Superatoms: Entanglement Generation and Transfer}}} (\bibinfo {year} {2025}),\ \Eprint {https://arxiv.org/abs/2504.12942} {arXiv:2504.12942} \BibitemShut {NoStop}%
\bibitem [{\citenamefont {Gustafsson}\ \emph {et~al.}(2014)\citenamefont {Gustafsson}, \citenamefont {Aref}, \citenamefont {Kockum}, \citenamefont {Ekstr{\"{o}}m}, \citenamefont {Johansson},\ and\ \citenamefont {Delsing}}]{Gustafsson2014}%
  \BibitemOpen
  \bibfield  {author} {\bibinfo {author} {\bibfnamefont {M.~V.}\ \bibnamefont {Gustafsson}}, \bibinfo {author} {\bibfnamefont {T.}~\bibnamefont {Aref}}, \bibinfo {author} {\bibfnamefont {A.~F.}\ \bibnamefont {Kockum}}, \bibinfo {author} {\bibfnamefont {M.~K.}\ \bibnamefont {Ekstr{\"{o}}m}}, \bibinfo {author} {\bibfnamefont {G.}~\bibnamefont {Johansson}},\ and\ \bibinfo {author} {\bibfnamefont {P.}~\bibnamefont {Delsing}},\ }\bibfield  {title} {\bibinfo {title} {{Propagating phonons coupled to an artificial atom}},\ }\href {https://doi.org/10.1126/science.1257219} {\bibfield  {journal} {\bibinfo  {journal} {Science}\ }\textbf {\bibinfo {volume} {346}},\ \bibinfo {pages} {207} (\bibinfo {year} {2014})}\BibitemShut {NoStop}%
\bibitem [{\citenamefont {Manenti}\ \emph {et~al.}(2017)\citenamefont {Manenti}, \citenamefont {Kockum}, \citenamefont {Patterson}, \citenamefont {Behrle}, \citenamefont {Rahamim}, \citenamefont {Tancredi}, \citenamefont {Nori},\ and\ \citenamefont {Leek}}]{Manenti2017}%
  \BibitemOpen
  \bibfield  {author} {\bibinfo {author} {\bibfnamefont {R.}~\bibnamefont {Manenti}}, \bibinfo {author} {\bibfnamefont {A.~F.}\ \bibnamefont {Kockum}}, \bibinfo {author} {\bibfnamefont {A.}~\bibnamefont {Patterson}}, \bibinfo {author} {\bibfnamefont {T.}~\bibnamefont {Behrle}}, \bibinfo {author} {\bibfnamefont {J.}~\bibnamefont {Rahamim}}, \bibinfo {author} {\bibfnamefont {G.}~\bibnamefont {Tancredi}}, \bibinfo {author} {\bibfnamefont {F.}~\bibnamefont {Nori}},\ and\ \bibinfo {author} {\bibfnamefont {P.~J.}\ \bibnamefont {Leek}},\ }\bibfield  {title} {\bibinfo {title} {{Circuit quantum acoustodynamics with surface acoustic waves}},\ }\href {https://doi.org/10.1038/s41467-017-01063-9} {\bibfield  {journal} {\bibinfo  {journal} {Nature Communications}\ }\textbf {\bibinfo {volume} {8}},\ \bibinfo {pages} {975} (\bibinfo {year} {2017})}\BibitemShut {NoStop}%
\bibitem [{\citenamefont {Satzinger}\ \emph {et~al.}(2018)\citenamefont {Satzinger}, \citenamefont {Zhong}, \citenamefont {Chang}, \citenamefont {Peairs}, \citenamefont {Bienfait}, \citenamefont {Chou}, \citenamefont {Cleland}, \citenamefont {Conner}, \citenamefont {Dumur}, \citenamefont {Grebel}, \citenamefont {Gutierrez}, \citenamefont {November}, \citenamefont {Povey}, \citenamefont {Whiteley}, \citenamefont {Awschalom}, \citenamefont {Schuster},\ and\ \citenamefont {Cleland}}]{Satzinger2018}%
  \BibitemOpen
  \bibfield  {author} {\bibinfo {author} {\bibfnamefont {K.~J.}\ \bibnamefont {Satzinger}}, \bibinfo {author} {\bibfnamefont {Y.~P.}\ \bibnamefont {Zhong}}, \bibinfo {author} {\bibfnamefont {H.-S.}\ \bibnamefont {Chang}}, \bibinfo {author} {\bibfnamefont {G.~A.}\ \bibnamefont {Peairs}}, \bibinfo {author} {\bibfnamefont {A.}~\bibnamefont {Bienfait}}, \bibinfo {author} {\bibfnamefont {M.-H.}\ \bibnamefont {Chou}}, \bibinfo {author} {\bibfnamefont {A.~Y.}\ \bibnamefont {Cleland}}, \bibinfo {author} {\bibfnamefont {C.~R.}\ \bibnamefont {Conner}}, \bibinfo {author} {\bibfnamefont {{\'{E}}.}~\bibnamefont {Dumur}}, \bibinfo {author} {\bibfnamefont {J.}~\bibnamefont {Grebel}}, \bibinfo {author} {\bibfnamefont {I.}~\bibnamefont {Gutierrez}}, \bibinfo {author} {\bibfnamefont {B.~H.}\ \bibnamefont {November}}, \bibinfo {author} {\bibfnamefont {R.~G.}\ \bibnamefont {Povey}}, \bibinfo {author} {\bibfnamefont {S.~J.}\ \bibnamefont {Whiteley}}, \bibinfo {author} {\bibfnamefont {D.~D.}\ \bibnamefont {Awschalom}}, \bibinfo
  {author} {\bibfnamefont {D.~I.}\ \bibnamefont {Schuster}},\ and\ \bibinfo {author} {\bibfnamefont {A.~N.}\ \bibnamefont {Cleland}},\ }\bibfield  {title} {\bibinfo {title} {{Quantum control of surface acoustic-wave phonons}},\ }\href {https://doi.org/10.1038/s41586-018-0719-5} {\bibfield  {journal} {\bibinfo  {journal} {Nature}\ }\textbf {\bibinfo {volume} {563}},\ \bibinfo {pages} {661} (\bibinfo {year} {2018})}\BibitemShut {NoStop}%
\bibitem [{\citenamefont {Bienfait}\ \emph {et~al.}(2019)\citenamefont {Bienfait}, \citenamefont {Satzinger}, \citenamefont {Zhong}, \citenamefont {Chang}, \citenamefont {Chou}, \citenamefont {Conner}, \citenamefont {Dumur}, \citenamefont {Grebel}, \citenamefont {Peairs}, \citenamefont {Povey},\ and\ \citenamefont {Cleland}}]{Bienfait2019}%
  \BibitemOpen
  \bibfield  {author} {\bibinfo {author} {\bibfnamefont {A.}~\bibnamefont {Bienfait}}, \bibinfo {author} {\bibfnamefont {K.~J.}\ \bibnamefont {Satzinger}}, \bibinfo {author} {\bibfnamefont {Y.~P.}\ \bibnamefont {Zhong}}, \bibinfo {author} {\bibfnamefont {H.-S.}\ \bibnamefont {Chang}}, \bibinfo {author} {\bibfnamefont {M.-H.}\ \bibnamefont {Chou}}, \bibinfo {author} {\bibfnamefont {C.~R.}\ \bibnamefont {Conner}}, \bibinfo {author} {\bibfnamefont {{\'{E}}.}~\bibnamefont {Dumur}}, \bibinfo {author} {\bibfnamefont {J.}~\bibnamefont {Grebel}}, \bibinfo {author} {\bibfnamefont {G.~A.}\ \bibnamefont {Peairs}}, \bibinfo {author} {\bibfnamefont {R.~G.}\ \bibnamefont {Povey}},\ and\ \bibinfo {author} {\bibfnamefont {A.~N.}\ \bibnamefont {Cleland}},\ }\bibfield  {title} {\bibinfo {title} {{Phonon-mediated quantum state transfer and remote qubit entanglement}},\ }\href {https://doi.org/10.1126/science.aaw8415} {\bibfield  {journal} {\bibinfo  {journal} {Science}\ }\textbf {\bibinfo {volume} {364}},\ \bibinfo {pages}
  {368} (\bibinfo {year} {2019})}\BibitemShut {NoStop}%
\bibitem [{\citenamefont {Andersson}\ \emph {et~al.}(2019)\citenamefont {Andersson}, \citenamefont {Suri}, \citenamefont {Guo}, \citenamefont {Aref},\ and\ \citenamefont {Delsing}}]{Andersson2019}%
  \BibitemOpen
  \bibfield  {author} {\bibinfo {author} {\bibfnamefont {G.}~\bibnamefont {Andersson}}, \bibinfo {author} {\bibfnamefont {B.}~\bibnamefont {Suri}}, \bibinfo {author} {\bibfnamefont {L.}~\bibnamefont {Guo}}, \bibinfo {author} {\bibfnamefont {T.}~\bibnamefont {Aref}},\ and\ \bibinfo {author} {\bibfnamefont {P.}~\bibnamefont {Delsing}},\ }\bibfield  {title} {\bibinfo {title} {{Non-exponential decay of a giant artificial atom}},\ }\href {https://doi.org/10.1038/s41567-019-0605-6} {\bibfield  {journal} {\bibinfo  {journal} {Nature Physics}\ }\textbf {\bibinfo {volume} {15}},\ \bibinfo {pages} {1123} (\bibinfo {year} {2019})}\BibitemShut {NoStop}%
\bibitem [{\citenamefont {Kannan}\ \emph {et~al.}(2020)\citenamefont {Kannan}, \citenamefont {Ruckriegel}, \citenamefont {Campbell}, \citenamefont {Frisk~Kockum}, \citenamefont {Braum\"{u}ller}, \citenamefont {Kim}, \citenamefont {Kjaergaard}, \citenamefont {Krantz}, \citenamefont {Melville}, \citenamefont {Niedzielski}, \citenamefont {Veps\"{a}l\"{a}inen}, \citenamefont {Winik}, \citenamefont {Yoder}, \citenamefont {Nori}, \citenamefont {Orlando}, \citenamefont {Gustavsson},\ and\ \citenamefont {Oliver}}]{Kannan2020}%
  \BibitemOpen
  \bibfield  {author} {\bibinfo {author} {\bibfnamefont {B.}~\bibnamefont {Kannan}}, \bibinfo {author} {\bibfnamefont {M.~J.}\ \bibnamefont {Ruckriegel}}, \bibinfo {author} {\bibfnamefont {D.~L.}\ \bibnamefont {Campbell}}, \bibinfo {author} {\bibfnamefont {A.}~\bibnamefont {Frisk~Kockum}}, \bibinfo {author} {\bibfnamefont {J.}~\bibnamefont {Braum\"{u}ller}}, \bibinfo {author} {\bibfnamefont {D.~K.}\ \bibnamefont {Kim}}, \bibinfo {author} {\bibfnamefont {M.}~\bibnamefont {Kjaergaard}}, \bibinfo {author} {\bibfnamefont {P.}~\bibnamefont {Krantz}}, \bibinfo {author} {\bibfnamefont {A.}~\bibnamefont {Melville}}, \bibinfo {author} {\bibfnamefont {B.~M.}\ \bibnamefont {Niedzielski}}, \bibinfo {author} {\bibfnamefont {A.}~\bibnamefont {Veps\"{a}l\"{a}inen}}, \bibinfo {author} {\bibfnamefont {R.}~\bibnamefont {Winik}}, \bibinfo {author} {\bibfnamefont {J.~L.}\ \bibnamefont {Yoder}}, \bibinfo {author} {\bibfnamefont {F.}~\bibnamefont {Nori}}, \bibinfo {author} {\bibfnamefont {T.~P.}\ \bibnamefont {Orlando}}, \bibinfo
  {author} {\bibfnamefont {S.}~\bibnamefont {Gustavsson}},\ and\ \bibinfo {author} {\bibfnamefont {W.~D.}\ \bibnamefont {Oliver}},\ }\bibfield  {title} {\bibinfo {title} {Waveguide quantum electrodynamics with superconducting artificial giant atoms},\ }\href {https://doi.org/10.1038/s41586-020-2529-9} {\bibfield  {journal} {\bibinfo  {journal} {Nature}\ }\textbf {\bibinfo {volume} {583}},\ \bibinfo {pages} {775} (\bibinfo {year} {2020})}\BibitemShut {NoStop}%
\bibitem [{\citenamefont {Bienfait}\ \emph {et~al.}(2020)\citenamefont {Bienfait}, \citenamefont {Zhong}, \citenamefont {Chang}, \citenamefont {Chou}, \citenamefont {Conner}, \citenamefont {Dumur}, \citenamefont {Grebel}, \citenamefont {Peairs}, \citenamefont {Povey}, \citenamefont {Satzinger},\ and\ \citenamefont {Cleland}}]{Bienfait2020}%
  \BibitemOpen
  \bibfield  {author} {\bibinfo {author} {\bibfnamefont {A.}~\bibnamefont {Bienfait}}, \bibinfo {author} {\bibfnamefont {Y.~P.}\ \bibnamefont {Zhong}}, \bibinfo {author} {\bibfnamefont {H.-S.}\ \bibnamefont {Chang}}, \bibinfo {author} {\bibfnamefont {M.-H.}\ \bibnamefont {Chou}}, \bibinfo {author} {\bibfnamefont {C.~R.}\ \bibnamefont {Conner}}, \bibinfo {author} {\bibfnamefont {{\'{E}}.}~\bibnamefont {Dumur}}, \bibinfo {author} {\bibfnamefont {J.}~\bibnamefont {Grebel}}, \bibinfo {author} {\bibfnamefont {G.~A.}\ \bibnamefont {Peairs}}, \bibinfo {author} {\bibfnamefont {R.~G.}\ \bibnamefont {Povey}}, \bibinfo {author} {\bibfnamefont {K.~J.}\ \bibnamefont {Satzinger}},\ and\ \bibinfo {author} {\bibfnamefont {A.~N.}\ \bibnamefont {Cleland}},\ }\bibfield  {title} {\bibinfo {title} {{Quantum Erasure Using Entangled Surface Acoustic Phonons}},\ }\href {https://doi.org/10.1103/PhysRevX.10.021055} {\bibfield  {journal} {\bibinfo  {journal} {Physical Review X}\ }\textbf {\bibinfo {volume} {10}},\ \bibinfo {pages}
  {021055} (\bibinfo {year} {2020})}\BibitemShut {NoStop}%
\bibitem [{\citenamefont {Andersson}\ \emph {et~al.}(2020)\citenamefont {Andersson}, \citenamefont {Ekstr{\"{o}}m},\ and\ \citenamefont {Delsing}}]{Andersson2020}%
  \BibitemOpen
  \bibfield  {author} {\bibinfo {author} {\bibfnamefont {G.}~\bibnamefont {Andersson}}, \bibinfo {author} {\bibfnamefont {M.~K.}\ \bibnamefont {Ekstr{\"{o}}m}},\ and\ \bibinfo {author} {\bibfnamefont {P.}~\bibnamefont {Delsing}},\ }\bibfield  {title} {\bibinfo {title} {{Electromagnetically Induced Acoustic Transparency with a Superconducting Circuit}},\ }\href {https://doi.org/10.1103/PhysRevLett.124.240402} {\bibfield  {journal} {\bibinfo  {journal} {Physical Review Letters}\ }\textbf {\bibinfo {volume} {124}},\ \bibinfo {pages} {240402} (\bibinfo {year} {2020})}\BibitemShut {NoStop}%
\bibitem [{\citenamefont {Vadiraj}\ \emph {et~al.}(2021)\citenamefont {Vadiraj}, \citenamefont {Ask}, \citenamefont {McConkey}, \citenamefont {Nsanzineza}, \citenamefont {Chang}, \citenamefont {Kockum},\ and\ \citenamefont {Wilson}}]{Vadiraj2021}%
  \BibitemOpen
  \bibfield  {author} {\bibinfo {author} {\bibfnamefont {A.~M.}\ \bibnamefont {Vadiraj}}, \bibinfo {author} {\bibfnamefont {A.}~\bibnamefont {Ask}}, \bibinfo {author} {\bibfnamefont {T.~G.}\ \bibnamefont {McConkey}}, \bibinfo {author} {\bibfnamefont {I.}~\bibnamefont {Nsanzineza}}, \bibinfo {author} {\bibfnamefont {C.~W.~S.}\ \bibnamefont {Chang}}, \bibinfo {author} {\bibfnamefont {A.~F.}\ \bibnamefont {Kockum}},\ and\ \bibinfo {author} {\bibfnamefont {C.~M.}\ \bibnamefont {Wilson}},\ }\bibfield  {title} {\bibinfo {title} {{Engineering the level structure of a giant artificial atom in waveguide quantum electrodynamics}},\ }\href {https://doi.org/10.1103/PhysRevA.103.023710} {\bibfield  {journal} {\bibinfo  {journal} {Physical Review A}\ }\textbf {\bibinfo {volume} {103}},\ \bibinfo {pages} {023710} (\bibinfo {year} {2021})}\BibitemShut {NoStop}%
\bibitem [{\citenamefont {Wang}\ \emph {et~al.}(2022)\citenamefont {Wang}, \citenamefont {Wang}, \citenamefont {Yao}, \citenamefont {Shen}, \citenamefont {Wu}, \citenamefont {Qian}, \citenamefont {Li}, \citenamefont {Zhu},\ and\ \citenamefont {You}}]{Wang2022a}%
  \BibitemOpen
  \bibfield  {author} {\bibinfo {author} {\bibfnamefont {Z.~Q.}\ \bibnamefont {Wang}}, \bibinfo {author} {\bibfnamefont {Y.~P.}\ \bibnamefont {Wang}}, \bibinfo {author} {\bibfnamefont {J.}~\bibnamefont {Yao}}, \bibinfo {author} {\bibfnamefont {R.~C.}\ \bibnamefont {Shen}}, \bibinfo {author} {\bibfnamefont {W.~J.}\ \bibnamefont {Wu}}, \bibinfo {author} {\bibfnamefont {J.}~\bibnamefont {Qian}}, \bibinfo {author} {\bibfnamefont {J.}~\bibnamefont {Li}}, \bibinfo {author} {\bibfnamefont {S.~Y.}\ \bibnamefont {Zhu}},\ and\ \bibinfo {author} {\bibfnamefont {J.~Q.}\ \bibnamefont {You}},\ }\bibfield  {title} {\bibinfo {title} {{Giant spin ensembles in waveguide magnonics}},\ }\href {https://doi.org/10.1038/s41467-022-35174-9} {\bibfield  {journal} {\bibinfo  {journal} {Nature Communications}\ }\textbf {\bibinfo {volume} {13}},\ \bibinfo {pages} {7580} (\bibinfo {year} {2022})}\BibitemShut {NoStop}%
\bibitem [{\citenamefont {Joshi}\ \emph {et~al.}(2023)\citenamefont {Joshi}, \citenamefont {Yang},\ and\ \citenamefont {Mirhosseini}}]{Joshi2023}%
  \BibitemOpen
  \bibfield  {author} {\bibinfo {author} {\bibfnamefont {C.}~\bibnamefont {Joshi}}, \bibinfo {author} {\bibfnamefont {F.}~\bibnamefont {Yang}},\ and\ \bibinfo {author} {\bibfnamefont {M.}~\bibnamefont {Mirhosseini}},\ }\bibfield  {title} {\bibinfo {title} {{Resonance Fluorescence of a Chiral Artificial Atom}},\ }\href {https://doi.org/10.1103/PhysRevX.13.021039} {\bibfield  {journal} {\bibinfo  {journal} {Physical Review X}\ }\textbf {\bibinfo {volume} {13}},\ \bibinfo {pages} {021039} (\bibinfo {year} {2023})}\BibitemShut {NoStop}%
\bibitem [{\citenamefont {Hu}\ \emph {et~al.}(2024)\citenamefont {Hu}, \citenamefont {Li}, \citenamefont {Qie}, \citenamefont {Yin}, \citenamefont {Kockum}, \citenamefont {Nori},\ and\ \citenamefont {An}}]{Hu2024}%
  \BibitemOpen
  \bibfield  {author} {\bibinfo {author} {\bibfnamefont {J.}~\bibnamefont {Hu}}, \bibinfo {author} {\bibfnamefont {D.}~\bibnamefont {Li}}, \bibinfo {author} {\bibfnamefont {Y.}~\bibnamefont {Qie}}, \bibinfo {author} {\bibfnamefont {Z.}~\bibnamefont {Yin}}, \bibinfo {author} {\bibfnamefont {A.~F.}\ \bibnamefont {Kockum}}, \bibinfo {author} {\bibfnamefont {F.}~\bibnamefont {Nori}},\ and\ \bibinfo {author} {\bibfnamefont {S.}~\bibnamefont {An}},\ }\href@noop {} {\bibinfo {title} {{Engineering the Environment of a Superconducting Qubit with an Artificial Giant Atom}}} (\bibinfo {year} {2024}),\ \Eprint {https://arxiv.org/abs/2410.15377} {arXiv:2410.15377} \BibitemShut {NoStop}%
\bibitem [{\citenamefont {Almanakly}\ \emph {et~al.}(2025)\citenamefont {Almanakly}, \citenamefont {Yankelevich}, \citenamefont {Hays}, \citenamefont {Kannan}, \citenamefont {Assouly}, \citenamefont {Greene}, \citenamefont {Gingras}, \citenamefont {Niedzielski}, \citenamefont {Stickler}, \citenamefont {Schwartz}, \citenamefont {Serniak}, \citenamefont {Wang}, \citenamefont {Orlando}, \citenamefont {Gustavsson}, \citenamefont {Grover},\ and\ \citenamefont {Oliver}}]{Almanakly2025}%
  \BibitemOpen
  \bibfield  {author} {\bibinfo {author} {\bibfnamefont {A.}~\bibnamefont {Almanakly}}, \bibinfo {author} {\bibfnamefont {B.}~\bibnamefont {Yankelevich}}, \bibinfo {author} {\bibfnamefont {M.}~\bibnamefont {Hays}}, \bibinfo {author} {\bibfnamefont {B.}~\bibnamefont {Kannan}}, \bibinfo {author} {\bibfnamefont {R.}~\bibnamefont {Assouly}}, \bibinfo {author} {\bibfnamefont {A.}~\bibnamefont {Greene}}, \bibinfo {author} {\bibfnamefont {M.}~\bibnamefont {Gingras}}, \bibinfo {author} {\bibfnamefont {B.~M.}\ \bibnamefont {Niedzielski}}, \bibinfo {author} {\bibfnamefont {H.}~\bibnamefont {Stickler}}, \bibinfo {author} {\bibfnamefont {M.~E.}\ \bibnamefont {Schwartz}}, \bibinfo {author} {\bibfnamefont {K.}~\bibnamefont {Serniak}}, \bibinfo {author} {\bibfnamefont {J.~{\^{I}}.-j.}\ \bibnamefont {Wang}}, \bibinfo {author} {\bibfnamefont {T.~P.}\ \bibnamefont {Orlando}}, \bibinfo {author} {\bibfnamefont {S.}~\bibnamefont {Gustavsson}}, \bibinfo {author} {\bibfnamefont {J.~A.}\ \bibnamefont {Grover}},\ and\ \bibinfo
  {author} {\bibfnamefont {W.~D.}\ \bibnamefont {Oliver}},\ }\bibfield  {title} {\bibinfo {title} {{Deterministic remote entanglement using a chiral quantum interconnect}},\ }\href {https://doi.org/10.1038/s41567-025-02811-1} {\bibfield  {journal} {\bibinfo  {journal} {Nature Physics}\ }\textbf {\bibinfo {volume} {21}},\ \bibinfo {pages} {825} (\bibinfo {year} {2025})}\BibitemShut {NoStop}%
\bibitem [{\citenamefont {Jouanny}\ \emph {et~al.}(2025)\citenamefont {Jouanny}, \citenamefont {Peyruchat}, \citenamefont {Scigliuzzo}, \citenamefont {Mercurio}, \citenamefont {{Di Benedetto}}, \citenamefont {{De Bernardis}}, \citenamefont {Sbroggi{\`{o}}}, \citenamefont {Frasca}, \citenamefont {Savona}, \citenamefont {Ciccarello},\ and\ \citenamefont {Scarlino}}]{Jouanny2025}%
  \BibitemOpen
  \bibfield  {author} {\bibinfo {author} {\bibfnamefont {V.}~\bibnamefont {Jouanny}}, \bibinfo {author} {\bibfnamefont {L.}~\bibnamefont {Peyruchat}}, \bibinfo {author} {\bibfnamefont {M.}~\bibnamefont {Scigliuzzo}}, \bibinfo {author} {\bibfnamefont {A.}~\bibnamefont {Mercurio}}, \bibinfo {author} {\bibfnamefont {E.}~\bibnamefont {{Di Benedetto}}}, \bibinfo {author} {\bibfnamefont {D.}~\bibnamefont {{De Bernardis}}}, \bibinfo {author} {\bibfnamefont {D.}~\bibnamefont {Sbroggi{\`{o}}}}, \bibinfo {author} {\bibfnamefont {S.}~\bibnamefont {Frasca}}, \bibinfo {author} {\bibfnamefont {V.}~\bibnamefont {Savona}}, \bibinfo {author} {\bibfnamefont {F.}~\bibnamefont {Ciccarello}},\ and\ \bibinfo {author} {\bibfnamefont {P.}~\bibnamefont {Scarlino}},\ }\href@noop {} {\bibinfo {title} {{Superstrong Dynamics and Chiral Emission of a Giant Atom in a Structured Bath}}} (\bibinfo {year} {2025}),\ \Eprint {https://arxiv.org/abs/2509.01579} {arXiv:2509.01579} \BibitemShut {NoStop}%
\bibitem [{\citenamefont {Guo}\ \emph {et~al.}(2020{\natexlab{b}})\citenamefont {Guo}, \citenamefont {Wang}, \citenamefont {Purdy},\ and\ \citenamefont {Taylor}}]{PhysRevA.102.033706}%
  \BibitemOpen
  \bibfield  {author} {\bibinfo {author} {\bibfnamefont {S.}~\bibnamefont {Guo}}, \bibinfo {author} {\bibfnamefont {Y.}~\bibnamefont {Wang}}, \bibinfo {author} {\bibfnamefont {T.}~\bibnamefont {Purdy}},\ and\ \bibinfo {author} {\bibfnamefont {J.}~\bibnamefont {Taylor}},\ }\bibfield  {title} {\bibinfo {title} {Beyond spontaneous emission: Giant atom bounded in the continuum},\ }\href {https://doi.org/10.1103/PhysRevA.102.033706} {\bibfield  {journal} {\bibinfo  {journal} {Physical Review A}\ }\textbf {\bibinfo {volume} {102}},\ \bibinfo {pages} {033706} (\bibinfo {year} {2020}{\natexlab{b}})}\BibitemShut {NoStop}%
\bibitem [{\citenamefont {Winkler}\ \emph {et~al.}(2006)\citenamefont {Winkler}, \citenamefont {Thalhammer}, \citenamefont {Lang}, \citenamefont {Grimm}, \citenamefont {Hecker~Denschlag}, \citenamefont {Daley}, \citenamefont {Kantian}, \citenamefont {B\"{u}chler},\ and\ \citenamefont {Zoller}}]{Winkler2006}%
  \BibitemOpen
  \bibfield  {author} {\bibinfo {author} {\bibfnamefont {K.}~\bibnamefont {Winkler}}, \bibinfo {author} {\bibfnamefont {G.}~\bibnamefont {Thalhammer}}, \bibinfo {author} {\bibfnamefont {F.}~\bibnamefont {Lang}}, \bibinfo {author} {\bibfnamefont {R.}~\bibnamefont {Grimm}}, \bibinfo {author} {\bibfnamefont {J.}~\bibnamefont {Hecker~Denschlag}}, \bibinfo {author} {\bibfnamefont {A.~J.}\ \bibnamefont {Daley}}, \bibinfo {author} {\bibfnamefont {A.}~\bibnamefont {Kantian}}, \bibinfo {author} {\bibfnamefont {H.~P.}\ \bibnamefont {B\"{u}chler}},\ and\ \bibinfo {author} {\bibfnamefont {P.}~\bibnamefont {Zoller}},\ }\bibfield  {title} {\bibinfo {title} {Repulsively bound atom pairs in an optical lattice},\ }\href {https://doi.org/10.1038/nature04918} {\bibfield  {journal} {\bibinfo  {journal} {Nature}\ }\textbf {\bibinfo {volume} {441}},\ \bibinfo {pages} {853} (\bibinfo {year} {2006})}\BibitemShut {NoStop}%
\bibitem [{\citenamefont {Piil}\ and\ \citenamefont {M\o{}lmer}(2007)}]{PhysRevA.76.023607}%
  \BibitemOpen
  \bibfield  {author} {\bibinfo {author} {\bibfnamefont {R.}~\bibnamefont {Piil}}\ and\ \bibinfo {author} {\bibfnamefont {K.}~\bibnamefont {M\o{}lmer}},\ }\bibfield  {title} {\bibinfo {title} {Tunneling couplings in discrete lattices, single-particle band structure, and eigenstates of interacting atom pairs},\ }\href {https://doi.org/10.1103/PhysRevA.76.023607} {\bibfield  {journal} {\bibinfo  {journal} {Physical Review A}\ }\textbf {\bibinfo {volume} {76}},\ \bibinfo {pages} {023607} (\bibinfo {year} {2007})}\BibitemShut {NoStop}%
\bibitem [{\citenamefont {Wang}\ \emph {et~al.}(2020)\citenamefont {Wang}, \citenamefont {Jaako}, \citenamefont {Kirton},\ and\ \citenamefont {Rabl}}]{PhysRevLett.124.213601}%
  \BibitemOpen
  \bibfield  {author} {\bibinfo {author} {\bibfnamefont {Z.}~\bibnamefont {Wang}}, \bibinfo {author} {\bibfnamefont {T.}~\bibnamefont {Jaako}}, \bibinfo {author} {\bibfnamefont {P.}~\bibnamefont {Kirton}},\ and\ \bibinfo {author} {\bibfnamefont {P.}~\bibnamefont {Rabl}},\ }\bibfield  {title} {\bibinfo {title} {{Supercorrelated Radiance in Nonlinear Photonic Waveguides}},\ }\href {https://doi.org/10.1103/PhysRevLett.124.213601} {\bibfield  {journal} {\bibinfo  {journal} {Physical Review Letters}\ }\textbf {\bibinfo {volume} {124}},\ \bibinfo {pages} {213601} (\bibinfo {year} {2020})}\BibitemShut {NoStop}%
\bibitem [{\citenamefont {Gorlach}\ and\ \citenamefont {Poddubny}(2017)}]{PhysRevA.95.053866}%
  \BibitemOpen
  \bibfield  {author} {\bibinfo {author} {\bibfnamefont {M.~A.}\ \bibnamefont {Gorlach}}\ and\ \bibinfo {author} {\bibfnamefont {A.~N.}\ \bibnamefont {Poddubny}},\ }\bibfield  {title} {\bibinfo {title} {Topological edge states of bound photon pairs},\ }\href {https://doi.org/10.1103/PhysRevA.95.053866} {\bibfield  {journal} {\bibinfo  {journal} {Physical Review A}\ }\textbf {\bibinfo {volume} {95}},\ \bibinfo {pages} {053866} (\bibinfo {year} {2017})}\BibitemShut {NoStop}%
\bibitem [{\citenamefont {Ayyash}\ \emph {et~al.}(2024)\citenamefont {Ayyash}, \citenamefont {Xu}, \citenamefont {Ashhab},\ and\ \citenamefont {Mariantoni}}]{PhysRevA.110.053711}%
  \BibitemOpen
  \bibfield  {author} {\bibinfo {author} {\bibfnamefont {M.}~\bibnamefont {Ayyash}}, \bibinfo {author} {\bibfnamefont {X.}~\bibnamefont {Xu}}, \bibinfo {author} {\bibfnamefont {S.}~\bibnamefont {Ashhab}},\ and\ \bibinfo {author} {\bibfnamefont {M.}~\bibnamefont {Mariantoni}},\ }\bibfield  {title} {\bibinfo {title} {Driven multiphoton qubit-resonator interactions},\ }\href {https://doi.org/10.1103/PhysRevA.110.053711} {\bibfield  {journal} {\bibinfo  {journal} {Physical Review A}\ }\textbf {\bibinfo {volume} {110}},\ \bibinfo {pages} {053711} (\bibinfo {year} {2024})}\BibitemShut {NoStop}%
\bibitem [{\citenamefont {Wegner}(1980)}]{wegner1980inverse}%
  \BibitemOpen
  \bibfield  {author} {\bibinfo {author} {\bibfnamefont {F.}~\bibnamefont {Wegner}},\ }\bibfield  {title} {\bibinfo {title} {Inverse participation ratio in 2+\ensuremath{\varepsilon} dimensions},\ }\href {https://doi.org/10.1007/BF01325284} {\bibfield  {journal} {\bibinfo  {journal} {Zeitschrift für Physik B Condensed Matter}\ }\textbf {\bibinfo {volume} {36}},\ \bibinfo {pages} {209} (\bibinfo {year} {1980})}\BibitemShut {NoStop}%
\bibitem [{foo()}]{footnote1}%
  \BibitemOpen
  \href@noop {} {}\bibinfo {howpublished} {See Supplemental Material at [url], which includes Refs.~\cite{wegner1980inverse,wang2024nonlinearchiralquantumoptics}, for additional information on the (i) numerical details and (ii) an analytical derivation of the effective model.}\BibitemShut {Stop}%
\bibitem [{\citenamefont {Andersson}\ \emph {et~al.}(2025)\citenamefont {Andersson}, \citenamefont {Havir}, \citenamefont {Ranni}, \citenamefont {Haldar},\ and\ \citenamefont {Maisi}}]{Andersson2025}%
  \BibitemOpen
  \bibfield  {author} {\bibinfo {author} {\bibfnamefont {S.}~\bibnamefont {Andersson}}, \bibinfo {author} {\bibfnamefont {H.}~\bibnamefont {Havir}}, \bibinfo {author} {\bibfnamefont {A.}~\bibnamefont {Ranni}}, \bibinfo {author} {\bibfnamefont {S.}~\bibnamefont {Haldar}},\ and\ \bibinfo {author} {\bibfnamefont {V.~F.}\ \bibnamefont {Maisi}},\ }\bibfield  {title} {\bibinfo {title} {High-impedance microwave resonators with two-photon nonlinear effects},\ }\href {https://doi.org/10.1038/s41467-025-55860-8} {\bibfield  {journal} {\bibinfo  {journal} {Nature Communications}\ }\textbf {\bibinfo {volume} {16}},\ \bibinfo {pages} {552} (\bibinfo {year} {2025})}\BibitemShut {NoStop}%
\bibitem [{\citenamefont {Gorshkov}\ \emph {et~al.}(2011)\citenamefont {Gorshkov}, \citenamefont {Otterbach}, \citenamefont {Fleischhauer}, \citenamefont {Pohl},\ and\ \citenamefont {Lukin}}]{PhysRevLett.107.133602}%
  \BibitemOpen
  \bibfield  {author} {\bibinfo {author} {\bibfnamefont {A.~V.}\ \bibnamefont {Gorshkov}}, \bibinfo {author} {\bibfnamefont {J.}~\bibnamefont {Otterbach}}, \bibinfo {author} {\bibfnamefont {M.}~\bibnamefont {Fleischhauer}}, \bibinfo {author} {\bibfnamefont {T.}~\bibnamefont {Pohl}},\ and\ \bibinfo {author} {\bibfnamefont {M.~D.}\ \bibnamefont {Lukin}},\ }\bibfield  {title} {\bibinfo {title} {{Photon-Photon Interactions via Rydberg Blockade}},\ }\href {https://doi.org/10.1103/PhysRevLett.107.133602} {\bibfield  {journal} {\bibinfo  {journal} {Physical Review Letters}\ }\textbf {\bibinfo {volume} {107}},\ \bibinfo {pages} {133602} (\bibinfo {year} {2011})}\BibitemShut {NoStop}%
\bibitem [{\citenamefont {Peyronel}\ \emph {et~al.}(2012)\citenamefont {Peyronel}, \citenamefont {Firstenberg}, \citenamefont {Liang}, \citenamefont {Hofferberth}, \citenamefont {Gorshkov}, \citenamefont {Pohl}, \citenamefont {Lukin},\ and\ \citenamefont {Vuletić}}]{Peyronel2012}%
  \BibitemOpen
  \bibfield  {author} {\bibinfo {author} {\bibfnamefont {T.}~\bibnamefont {Peyronel}}, \bibinfo {author} {\bibfnamefont {O.}~\bibnamefont {Firstenberg}}, \bibinfo {author} {\bibfnamefont {Q.-Y.}\ \bibnamefont {Liang}}, \bibinfo {author} {\bibfnamefont {S.}~\bibnamefont {Hofferberth}}, \bibinfo {author} {\bibfnamefont {A.~V.}\ \bibnamefont {Gorshkov}}, \bibinfo {author} {\bibfnamefont {T.}~\bibnamefont {Pohl}}, \bibinfo {author} {\bibfnamefont {M.~D.}\ \bibnamefont {Lukin}},\ and\ \bibinfo {author} {\bibfnamefont {V.}~\bibnamefont {Vuletić}},\ }\bibfield  {title} {\bibinfo {title} {Quantum nonlinear optics with single photons enabled by strongly interacting atoms},\ }\href {https://doi.org/10.1038/nature11361} {\bibfield  {journal} {\bibinfo  {journal} {Nature}\ }\textbf {\bibinfo {volume} {488}},\ \bibinfo {pages} {57} (\bibinfo {year} {2012})}\BibitemShut {NoStop}%
\bibitem [{\citenamefont {Chen}\ and\ \citenamefont {Frisk~Kockum}(2025)}]{Chen2025}%
  \BibitemOpen
  \bibfield  {author} {\bibinfo {author} {\bibfnamefont {G.}~\bibnamefont {Chen}}\ and\ \bibinfo {author} {\bibfnamefont {A.}~\bibnamefont {Frisk~Kockum}},\ }\bibfield  {title} {\bibinfo {title} {Simulating open quantum systems with giant atoms},\ }\href {https://doi.org/10.1088/2058-9565/adb2bd} {\bibfield  {journal} {\bibinfo  {journal} {Quantum Science and Technology}\ }\textbf {\bibinfo {volume} {10}},\ \bibinfo {pages} {025028} (\bibinfo {year} {2025})}\BibitemShut {NoStop}%
\bibitem [{\citenamefont {Chen}\ and\ \citenamefont {Kockum}(2025{\natexlab{a}})}]{chen2025scalablequantumsimulatorextended}%
  \BibitemOpen
  \bibfield  {author} {\bibinfo {author} {\bibfnamefont {G.}~\bibnamefont {Chen}}\ and\ \bibinfo {author} {\bibfnamefont {A.~F.}\ \bibnamefont {Kockum}},\ }\href@noop {} {\bibinfo {title} {Scalable quantum simulator with an extended gate set in giant atoms}} (\bibinfo {year} {2025}{\natexlab{a}}),\ \Eprint {https://arxiv.org/abs/2503.04537} {arXiv:2503.04537} \BibitemShut {NoStop}%
\bibitem [{\citenamefont {Chen}\ and\ \citenamefont {Kockum}(2025{\natexlab{b}})}]{Chen2025b}%
  \BibitemOpen
  \bibfield  {author} {\bibinfo {author} {\bibfnamefont {G.}~\bibnamefont {Chen}}\ and\ \bibinfo {author} {\bibfnamefont {A.~F.}\ \bibnamefont {Kockum}},\ }\href@noop {} {\bibinfo {title} {{Efficient three-qubit gates with giant atoms}}} (\bibinfo {year} {2025}{\natexlab{b}}),\ \Eprint {https://arxiv.org/abs/2510.04545} {arXiv:2510.04545} \BibitemShut {NoStop}%
\bibitem [{\citenamefont {D\"ur}\ \emph {et~al.}(2003)\citenamefont {D\"ur}, \citenamefont {Aschauer},\ and\ \citenamefont {Briegel}}]{PhysRevLett.91.107903}%
  \BibitemOpen
  \bibfield  {author} {\bibinfo {author} {\bibfnamefont {W.}~\bibnamefont {D\"ur}}, \bibinfo {author} {\bibfnamefont {H.}~\bibnamefont {Aschauer}},\ and\ \bibinfo {author} {\bibfnamefont {H.-J.}\ \bibnamefont {Briegel}},\ }\bibfield  {title} {\bibinfo {title} {{Multiparticle Entanglement Purification for Graph States}},\ }\href {https://doi.org/10.1103/PhysRevLett.91.107903} {\bibfield  {journal} {\bibinfo  {journal} {Physical Review Letters}\ }\textbf {\bibinfo {volume} {91}},\ \bibinfo {pages} {107903} (\bibinfo {year} {2003})}\BibitemShut {NoStop}%
\bibitem [{\citenamefont {Briegel}\ \emph {et~al.}(1998)\citenamefont {Briegel}, \citenamefont {D\"ur}, \citenamefont {Cirac},\ and\ \citenamefont {Zoller}}]{PhysRevLett.81.5932}%
  \BibitemOpen
  \bibfield  {author} {\bibinfo {author} {\bibfnamefont {H.-J.}\ \bibnamefont {Briegel}}, \bibinfo {author} {\bibfnamefont {W.}~\bibnamefont {D\"ur}}, \bibinfo {author} {\bibfnamefont {J.~I.}\ \bibnamefont {Cirac}},\ and\ \bibinfo {author} {\bibfnamefont {P.}~\bibnamefont {Zoller}},\ }\bibfield  {title} {\bibinfo {title} {{Quantum Repeaters: The Role of Imperfect Local Operations in Quantum Communication}},\ }\href {https://doi.org/10.1103/PhysRevLett.81.5932} {\bibfield  {journal} {\bibinfo  {journal} {Physical Review Letters}\ }\textbf {\bibinfo {volume} {81}},\ \bibinfo {pages} {5932} (\bibinfo {year} {1998})}\BibitemShut {NoStop}%
\bibitem [{\citenamefont {Wehner}\ \emph {et~al.}(2018)\citenamefont {Wehner}, \citenamefont {Elkouss},\ and\ \citenamefont {Hanson}}]{Wehner2018}%
  \BibitemOpen
  \bibfield  {author} {\bibinfo {author} {\bibfnamefont {S.}~\bibnamefont {Wehner}}, \bibinfo {author} {\bibfnamefont {D.}~\bibnamefont {Elkouss}},\ and\ \bibinfo {author} {\bibfnamefont {R.}~\bibnamefont {Hanson}},\ }\bibfield  {title} {\bibinfo {title} {Quantum internet: A vision for the road ahead},\ }\href {https://doi.org/10.1126/science.aam9288} {\bibfield  {journal} {\bibinfo  {journal} {Science}\ }\textbf {\bibinfo {volume} {362}},\ \bibinfo {pages} {eaam9288} (\bibinfo {year} {2018})}\BibitemShut {NoStop}%
\bibitem [{\citenamefont {Sangouard}\ \emph {et~al.}(2011)\citenamefont {Sangouard}, \citenamefont {Simon}, \citenamefont {de~Riedmatten},\ and\ \citenamefont {Gisin}}]{RevModPhys.83.33}%
  \BibitemOpen
  \bibfield  {author} {\bibinfo {author} {\bibfnamefont {N.}~\bibnamefont {Sangouard}}, \bibinfo {author} {\bibfnamefont {C.}~\bibnamefont {Simon}}, \bibinfo {author} {\bibfnamefont {H.}~\bibnamefont {de~Riedmatten}},\ and\ \bibinfo {author} {\bibfnamefont {N.}~\bibnamefont {Gisin}},\ }\bibfield  {title} {\bibinfo {title} {Quantum repeaters based on atomic ensembles and linear optics},\ }\href {https://doi.org/10.1103/RevModPhys.83.33} {\bibfield  {journal} {\bibinfo  {journal} {Reviews of Modern Physics}\ }\textbf {\bibinfo {volume} {83}},\ \bibinfo {pages} {33} (\bibinfo {year} {2011})}\BibitemShut {NoStop}%
\bibitem [{\citenamefont {Caneva}\ \emph {et~al.}(2015)\citenamefont {Caneva}, \citenamefont {Manzoni}, \citenamefont {Shi}, \citenamefont {Douglas}, \citenamefont {Cirac},\ and\ \citenamefont {Chang}}]{Caneva2015}%
  \BibitemOpen
  \bibfield  {author} {\bibinfo {author} {\bibfnamefont {T.}~\bibnamefont {Caneva}}, \bibinfo {author} {\bibfnamefont {M.~T.}\ \bibnamefont {Manzoni}}, \bibinfo {author} {\bibfnamefont {T.}~\bibnamefont {Shi}}, \bibinfo {author} {\bibfnamefont {J.~S.}\ \bibnamefont {Douglas}}, \bibinfo {author} {\bibfnamefont {J.~I.}\ \bibnamefont {Cirac}},\ and\ \bibinfo {author} {\bibfnamefont {D.~E.}\ \bibnamefont {Chang}},\ }\bibfield  {title} {\bibinfo {title} {Quantum dynamics of propagating photons with strong interactions: a generalized input–output formalism},\ }\href {https://doi.org/10.1088/1367-2630/17/11/113001} {\bibfield  {journal} {\bibinfo  {journal} {New Journal of Physics}\ }\textbf {\bibinfo {volume} {17}},\ \bibinfo {pages} {113001} (\bibinfo {year} {2015})}\BibitemShut {NoStop}%
\bibitem [{\citenamefont {Kimble}(2008)}]{Kimble2008}%
  \BibitemOpen
  \bibfield  {author} {\bibinfo {author} {\bibfnamefont {H.~J.}\ \bibnamefont {Kimble}},\ }\bibfield  {title} {\bibinfo {title} {The quantum internet},\ }\href {https://doi.org/10.1038/nature07127} {\bibfield  {journal} {\bibinfo  {journal} {Nature}\ }\textbf {\bibinfo {volume} {453}},\ \bibinfo {pages} {1023} (\bibinfo {year} {2008})}\BibitemShut {NoStop}%
\bibitem [{\citenamefont {Papanicolaou}\ and\ \citenamefont {Psaltakis}(1987)}]{PhysRevB.35.342}%
  \BibitemOpen
  \bibfield  {author} {\bibinfo {author} {\bibfnamefont {N.}~\bibnamefont {Papanicolaou}}\ and\ \bibinfo {author} {\bibfnamefont {G.~C.}\ \bibnamefont {Psaltakis}},\ }\bibfield  {title} {\bibinfo {title} {Bethe ansatz for two-magnon bound states in anisotropic magnetic chains of arbitrary spin},\ }\href {https://doi.org/10.1103/PhysRevB.35.342} {\bibfield  {journal} {\bibinfo  {journal} {Physical Review B}\ }\textbf {\bibinfo {volume} {35}},\ \bibinfo {pages} {342} (\bibinfo {year} {1987})}\BibitemShut {NoStop}%
\bibitem [{\citenamefont {Wu}\ \emph {et~al.}(2022)\citenamefont {Wu}, \citenamefont {Katsura}, \citenamefont {Li}, \citenamefont {Cai},\ and\ \citenamefont {Guan}}]{PhysRevB.105.064419}%
  \BibitemOpen
  \bibfield  {author} {\bibinfo {author} {\bibfnamefont {N.}~\bibnamefont {Wu}}, \bibinfo {author} {\bibfnamefont {H.}~\bibnamefont {Katsura}}, \bibinfo {author} {\bibfnamefont {S.-W.}\ \bibnamefont {Li}}, \bibinfo {author} {\bibfnamefont {X.}~\bibnamefont {Cai}},\ and\ \bibinfo {author} {\bibfnamefont {X.-W.}\ \bibnamefont {Guan}},\ }\bibfield  {title} {\bibinfo {title} {{Exact solutions of few-magnon problems in the spin-$S$ periodic XXZ chain}},\ }\href {https://doi.org/10.1103/PhysRevB.105.064419} {\bibfield  {journal} {\bibinfo  {journal} {Physical Review B}\ }\textbf {\bibinfo {volume} {105}},\ \bibinfo {pages} {064419} (\bibinfo {year} {2022})}\BibitemShut {NoStop}%
\bibitem [{\citenamefont {Takahashi}(1999)}]{Takahashi1999}%
  \BibitemOpen
  \bibfield  {author} {\bibinfo {author} {\bibfnamefont {M.}~\bibnamefont {Takahashi}},\ }\href {https://doi.org/10.1017/cbo9780511524332} {\emph {\bibinfo {title} {{Thermodynamics of One-Dimensional Solvable Models}}}}\ (\bibinfo  {publisher} {Cambridge University Press},\ \bibinfo {year} {1999})\BibitemShut {NoStop}%
\bibitem [{\citenamefont {Essler}\ \emph {et~al.}(2005)\citenamefont {Essler}, \citenamefont {Frahm}, \citenamefont {G\"{o}hmann}, \citenamefont {Kl\"{u}mper},\ and\ \citenamefont {Korepin}}]{Essler2005}%
  \BibitemOpen
  \bibfield  {author} {\bibinfo {author} {\bibfnamefont {F.~H.~L.}\ \bibnamefont {Essler}}, \bibinfo {author} {\bibfnamefont {H.}~\bibnamefont {Frahm}}, \bibinfo {author} {\bibfnamefont {F.}~\bibnamefont {G\"{o}hmann}}, \bibinfo {author} {\bibfnamefont {A.}~\bibnamefont {Kl\"{u}mper}},\ and\ \bibinfo {author} {\bibfnamefont {V.~E.}\ \bibnamefont {Korepin}},\ }\href {https://doi.org/10.1017/cbo9780511534843} {\emph {\bibinfo {title} {{The One-Dimensional Hubbard Model}}}}\ (\bibinfo  {publisher} {Cambridge University Press},\ \bibinfo {year} {2005})\BibitemShut {NoStop}%
\bibitem [{\citenamefont {Wang}\ \emph {et~al.}(2024{\natexlab{b}})\citenamefont {Wang}, \citenamefont {Li}, \citenamefont {Wang}, \citenamefont {Kockum}, \citenamefont {Du}, \citenamefont {Liu},\ and\ \citenamefont {Nori}}]{wang2024nonlinearchiralquantumoptics}%
  \BibitemOpen
  \bibfield  {author} {\bibinfo {author} {\bibfnamefont {X.}~\bibnamefont {Wang}}, \bibinfo {author} {\bibfnamefont {J.-Q.}\ \bibnamefont {Li}}, \bibinfo {author} {\bibfnamefont {Z.}~\bibnamefont {Wang}}, \bibinfo {author} {\bibfnamefont {A.~F.}\ \bibnamefont {Kockum}}, \bibinfo {author} {\bibfnamefont {L.}~\bibnamefont {Du}}, \bibinfo {author} {\bibfnamefont {T.}~\bibnamefont {Liu}},\ and\ \bibinfo {author} {\bibfnamefont {F.}~\bibnamefont {Nori}},\ }\href@noop {} {\bibinfo {title} {Nonlinear chiral quantum optics with giant-emitter pairs}} (\bibinfo {year} {2024}{\natexlab{b}}),\ \Eprint {https://arxiv.org/abs/2404.09829} {arXiv:2404.09829} \BibitemShut {NoStop}%
\end{thebibliography}
\end{document}


\author{Walter Rieck}
\affiliation{Department of Microtechnology and Nanoscience, Chalmers University of Technology, 41296 Gothenburg, Sweden}


\author{Anton Frisk Kockum}
\email{anton.frisk.kockum@chalmers.se}
\affiliation{Department of Microtechnology and Nanoscience, Chalmers University of Technology, 41296 Gothenburg, Sweden}

\author{Guangze Chen}
\email{guangze@chalmers.se}
\affiliation{Department of Microtechnology and Nanoscience, Chalmers University of Technology, 41296 Gothenburg, Sweden}

\title{Supplemental Material for ``Doublon bound states in the continuum through giant atoms''}

\date{\today}

\maketitle


\section{Numerical details}

In the main text, we defined the full Hamiltonian of the system ($N_a$ giant atoms interacting with a waveguide consisting of $N$ cavities at $p_n$ coupling points for atom $n$) in Eqs.~(1), (2), (4), and (9) (for the excitation-number-conserving interaction Hamiltonian $H_{\rm int}$):
\begin{align}
    H &= H_A + H_B + H_{\rm int} , \\
    H_A &= \sum_{n=1}^{N_a} \mleft[ {\mleft(\Delta_2 + \Delta_1\mright)} \frac{\sigma_n^{(2), +} \sigma_n^{(2), -}}{2} + \Delta_1 \sigma_n^{(1), +} \sigma_n^{(1), -} \mright], \label{eq:HA-general}\\
    H_B &= \sum_{n = 1}^N \frac{U}{2}a_n^{\dag}a_n^{\dag}a_n a_n - \sum_{n = 1}^{N-1}J (a_n^{\dag}a_{n+1} + \text{H.c.}) , \label{eq:HB-general} \\
    H_{\rm int} &= g\sum_{n=1}^{N_a}\sum_{j=1}^{p_n} \mleft[\sigma_n^{(1), +}a_{x_{nj}}+\sigma_n^{(2), +}a_{x_{nj}} + \text{H.c.} \mright]. \label{eq:Hint-general-number-preserving}
\end{align}

To perform numerical simulations for time evolution of the system with $N_a = 2$ giant atoms under this Hamiltonian, we first identify a complete basis for the system in the two-excitation subspace by enumerating all physically distinct states associated with each number of atomic excitations in this subspace:
\begin{enumerate}
\item For two atomic excitations and no photons, there are three distinct states: $\ket{02, 00}$, $\ket{20, 00}$, and $\ket{11, 00}$.
\item For one atomic excitation and one photon in the waveguide, there are $2N$ states: $\ket{10, n0}$ and $\ket{01, n0}$ for each mode index $n$.
\item For no atomic excitations and two photons in the waveguide, we have $N(N+1)/2$ states of the form $\ket{00, nm}$, where $\ket{00, nm} \equiv \ket{00, mn}$ enforces symmetry under exchange.
\end{enumerate}
The basis states here are explicitly defined as
\begin{align}
    &\ket{20, 00} = \frac{\sigma_1^{(2), +}\sigma_1^{(1), +}}{\sqrt{2}}\ket{\text{vac}}, \;\;  
    \ket{02, 00} = \frac{\sigma_2^{(2), +}\sigma_2^{(1), +}}{\sqrt{2}}\ket{\text{vac}}, \nonumber \\
    &\ket{11, 00} = \sigma_1^{(1), +}\sigma_2^{(1), +}\ket{\text{vac}}, \;\;
    \ket{10, n0} = \sigma_1^{(1), +}a_n^{\dag}\ket{\text{vac}}, \nonumber \\
    &\ket{01, n0} = \sigma_2^{(1), +}a_n^{\dag}\ket{\text{vac}},\quad \;\:\:\:
    \ket{00, nm} = a_n^{\dag}a_m^{\dag}\ket{\text{vac}}. 
\end{align}

Matrix elements of the Hamiltonian are computed by evaluating inner products between basis states and the Hamiltonian operator. For example, 
\beq
\begin{aligned}
    \bra{20, 00}H\ket{20, 00} = \bra{20, 00}H_A\ket{20, 00} = \Delta_2 + \Delta_1 .
\end{aligned}
\eeq

We use Algorithm~\ref{algo:ConstructHamiltonian} to construct the Hamiltonian matrix numerically. By diagonalizing the constructed Hamiltonian, we obtain its eigenvalues and vectors. We then use the inverse participation ratio (IPR) to identify highly localized states that potentially are bound states in the continuum. The IPR for a state $\ket{\psi}$ is defined as~\cite{wegner1980inverse}
\beq
\text{IPR}(\ket{\psi}) = \sum_n |\psi_n|^4,
\eeq
where $\psi_n$ is the amplitude of the $n$th basis state of the Hamiltonian. A value of $\text{IPR} \approx 1$ indicates a highly localized state, while $\text{IPR} \approx 1/\sqrt{N}$ suggests a delocalized state. The BICs can therefore be found by locating the eigenstates that have the highest IPR values. In systems where BICs exist, the eigenstate with either the largest or second largest IPR typically corresponds to the BIC.


\section{Analytical derivation}

Here we derive the effective Hamiltonian $H_\text{eff}$ for a single giant atom coupled at two points to the structured waveguide with the excitation-number-conserving interaction Hamiltonian $H_{\rm int}$ given in Eq.~(9) in the main text, such that Eqs.~(\ref{eq:HA-general})--(\ref{eq:Hint-general-number-preserving}) are simplified to
\beq
\begin{aligned}
H_A &=  \mleft[(\Delta_2+\Delta_1)\sigma^{(2), +}\sigma^{(2), -} + \Delta_1\sigma^{(1), +}\sigma^{(1), -}\mright]\\
H_B &= \sum_{n = 1}^N \frac{U}{2}a_n^{\dag}a_n^{\dag}a_n a_n - \sum_{n = 1}^{N-1}J (a_n^{\dag}a_{n+1} + \text{H.c.})\\
H_{\rm int} &= g\sum_{j=1}^{2} \mleft[\sigma^{(1), +}a_{x_j}+\sigma^{(2), +}a_{x_j} + \text{H.c.} \mright].
\end{aligned}
\eeq
where $x_{1,2}$ are the coupling-point coordinates of the atom. In particular, we are interested in the regime where the energy of the $\ket{2}$ level of the atom lies within the doublon band: $\Delta_1+\Delta_2\approx\omega_D(k)$. We therefore derive the effective Hamiltonian in the subspace spanned by the states $\ket{2}$ and $\ket{\psi_D(k)}$.

\onecolumngrid
\begin{algorithm}[t]
\caption{ConstructHamiltonian}
\label{algo:ConstructHamiltonian}

\SetKwComment{Comment}{\texttt{// }}{}
\SetKw{KwLet}{Let}
\SetKw{KwInit}{Initialize}
\KwLet $D = \binom{N + N_a + 1}{2}$\;
\KwInit symmetric Hamiltonian $H \in \mathbb{C}^{D \times D}$ with zeros, $g[i][j]$ as the coupling strength of atom $i$ at coupling point $j$, couplePoints[i] as the coupling point indices of atom i, and $\Delta_1[i]$ and $\Delta_2[i]$ as the level spacings associated with level $1$ and $2$ respectively. 

\Comment{Set energy levels and couplings for $\ket{2}$}
\For{$i = 0$ \KwTo $\binom{N_a + 1}{2} - 1$}{
    \If{$i < N_a$}{
        $H[i, i] = \Delta_1[i] + \Delta_2[i]$\;
        \ForEach{couplePoint, $k$ in couplePoints$[i]$}{
            $j = \binom{N_a + 1}{2} + i N + (\text{couplePoint} - 1)$\;
            $H[i, j] =\sqrt{2} g[i][k]$\;
        }
    }
    \Else{
        \ForEach{$0<j<k < N_a$}{
            \If{$i == jN - j(j-1)/2 + k - j$}{
                $H[i,i] = \Delta_1[j] + \Delta_1[k]$
            }
            
        }
    }
}

\For{$i = 0$ \KwTo $N_a - 2$}{
    \For{$j = i+1$ \KwTo $N_a - 1$}{
        $\text{index} = (i + 1)N_a - i(i - 1)/2 + j - 2i - 1$\;
        \ForEach{(\text{couplePoint}, $k$) in \text{couplePoints}$[i]$}{
            $H[\text{index}, \binom{N_a + 1}{2} + jN + \text{couplePoint} - 1] = g[i][k]$\;
        }
        \ForEach{(\text{couplePoint}, k) in \text{couplePoints}[j])}{
            $H[\text{index}, \binom{N_a + 1}{2} + jN + \text{couplePoint} - 1] = g[j][k]$\;
        }
    }
}

\Comment{Set energy levels for couplings for $\ket{1}$}
\For{$i = 0$ \KwTo $N_a - 1$}{
    \For{$j = 0$ \KwTo $N - 1$}{
        $\text{index} = \binom{N_a + 1}{2} + i N + j$\;
        $H[\text{index}, \text{index}] = \Delta_1[i]$\;
        \If{$j \neq N - 1$}{
            $H[\text{index}, \text{index} + 1] = -J$\;
        }
        iter $= 0$\;
        \ForEach{couplePoint in couplePoints$[i]$}{
            $k = \text{couplePoint} - 1$\;
            \If{$j \leq k$} {
                $c_k =  jN - j(j - 1)/2 + k - j$\;
            }
            \Else{
                $c_k = kN - k(k - 1)/2 + j - k$\;
            }
            $H[\text{index}, c_k + \binom{N_a + 1}{2} + NN_a] = g[i][\text{iter}]$
            
            \If{$j == k$}{
                $H[i,c_k+ \binom{N_a + 1}{2} + N N_a] = g[i][\text{iter}]$\;
                $H[\text{index},c_k+ \binom{N_a + 1}{2} + N N_a] = \sqrt{2} \cdot g[i][\text{iter}]$\;
            }
            iter++
        }
    }
}
\Comment{Set energy levels and coupling for $\ket{0}$}
\For{$i = 0$ \KwTo $\binom{N+1}{2} - 1$}{
    $\text{index} = i + \binom{N_a + 1}{2} + N_a N$\;
    \For{$k = 0$ \KwTo $N - 1$}{
        \If{$i < kN - k(k-1)/2$}{
            $c_k = i - N(k - 1) + \frac{(k-2)(k-1)}{2} + (k - 1)$, $c_j = k - 1$\;
            break;
        }
        \Else{
            $c_k = N - 1$, $c_j = N - 1$\;
        }
    }
    Determine neighbors of $(c_j, c_k)$: $(c_j \pm 1, c_k)$ and $(c_j, c_k \pm 1)$\;
    \ForEach{\text{valid neighbor} $(j', k')$}{
        $c_{jk} = Nj' - j'(j' - 1)/2 + k' - j'$\;
        \If{$j' = k'$}{
            $H[\text{index}, c_{jk}] = -\sqrt{2} J$\;
            $H[c_{jk}, c_{jk}] = U$\;
        }
        \Else{
            $H[\text{index}, c_{jk}] = -J$\;
        }
    }
}
\end{algorithm}
\twocolumngrid



Since $H_{\rm int}$ does not directly couple these states (i.e., $\bra{\psi_D(k)}H_{\rm int}\ket{2}=0$), the lowest non-zero contribution arises from second-order perturbation theory:
\beq
\bra{\psi_D(k)}H_\text{eff}\ket{2}=\sum_m \frac{\bra{\psi_D(k)}H_{\rm int}\ket{m}\bra{m}H_{\rm int}\ket{2}}{\Delta_1+\Delta_2-E_m}
\eeq
where $\ket{m}$ are eigenstates of $H_A+H_B$. To yield a non-zero contribution, both matrix elements in the numerator must be non-zero. The only such intermediate states $\ket{m}$ are $\ket{1,k'}=\sigma^{(1), +}a^\dag_{k'}\ket{\text{vac}}$ with $E(k')=\Delta_1+2J\cos k'$. 

We first Fourier transform the interaction Hamiltonian:
\beq
H_{\rm int} = \frac{g}{\sqrt{N}}\sum_{j=1}^{2}\sum_k \mleft[\sigma^{(1), +}+\sigma^{(2), +}\mright]\exp(ikx_j)a_k + \text{H.c.} ,
\eeq
which gives
\beq
\bra{1,k'}H_{\rm int}\ket{2}=\frac{\sqrt{2}g}{\sqrt{N}}\sum_{j=1}^{2}\exp(ik'x_j):=G(k').
\eeq

To evaluate $\bra{\psi_D(k)}H_{\rm int}\ket{1,k'}$, we note that:
\beq
\ket{\psi_D(k)}=\frac{1}{\sqrt{2}}\sum_{m=1}^N\sum_{n=1}^Ne^{ik\frac{(m+n)}{2}}e^{-\frac{|m-n|}{\varsigma(k)}}a^\dag_ma^\dag_n\ket{\text{vac}},
\eeq
where the localization length $\varsigma(k)$ is given by~\cite{wang2024nonlinearchiralquantumoptics}
\beq
\varsigma(k)=-\mleft[\ln\mleft(\frac{\sqrt{U^2+16J^2\cos(k/2)^2}-U}{4J\cos(k/2)}\mright)\mright]^{-1} .
\eeq
For large $U$, such as $U=10J$ considered in the main text, the localization length $\varsigma(k)$ is sufficiently small [e.g., $\varsigma(\pi/2)\approx0.506$] that we can approximate $\exp{\mleft[-\frac{|m-n|}{\varsigma(k)}\mright]}$ as $\delta_{mn}$, yielding
\beq
\ket{\psi_D(k)}=\frac{1}{\sqrt{2}}\sum_{n=1}^N\exp{\mleft(ikn\mright)}a^\dag_na^\dag_n\ket{\text{vac}} 
\eeq
and
\beq
\bra{\psi_D(k)}H_{\rm int}\ket{1,k'}=G^*(k-k').
\eeq

Therefore, the second-order matrix element becomes
\beq \label{eqS12}
\begin{aligned}
&\bra{\psi_D(k)}H_\text{eff}\ket{2}\\
=&\sum_{k'}\frac{1}{\Delta_2-2J\cos k'}G(k-k')G(k')\\
=&\frac{2g^2}{N}\sum_{k'}\frac{e^{ikx_1}+e^{ikx_2}}{\Delta_2-2J\cos k'}\\
&+\frac{2g^2}{N}\sum_{k'}\frac{e^{ikx_1}e^{ik'(x_2-x_1)}+e^{ikx_2}e^{ik'(x_1-x_2)}}{\Delta_2-2J\cos k'}.
\end{aligned}
\eeq
Due to the oscillation in $k'$, the second term is suppressed, and the first term in \eqref{eqS12} dominates. In particular, at the decoherence-free wavevector $k=\pi/(x_2-x_1)$, the second term vanishes. This yields
\beq \label{eqS14}
\bra{\psi_D(k)}H_\text{eff}\ket{2}= g_\text{eff}\sum_{j=1} ^{2}\exp(ikx_j) ,
\eeq
where
\beq
g_\text{eff}=\frac{2g^2}{N}\sum_{k'}\frac{1}{\Delta_2-2J\cos k'}.
\eeq
For the parameters used in Fig.~3(a–d) of the main text ($N=199$, $U = 10J$, $g = 0.25 J$, $\Delta_2=5J$, and $x_2-x_1 = 2$), this effective coupling takes a value of $g_\text{eff} \approx 0.027J$.

The expression \eqref{eqS14} mirrors the single-photon matrix element
\beq
\bra{\rm vac} a_k H_{\rm int} \ket{1} = g \sum_{j=1}^{2} \exp(ikx_j),
\eeq
except with $g\to g_\text{eff}$. The factor $\sum_{j=1}^{2}\exp(ikx_j)$, characteristic of giant atoms, gives rise to interference effects and underpins both bound states in the continuum (BICs) and decoherence-free interactions (DFIs).

In addition to the effective coupling, the second-order correction also induces a Lamb shift $\Delta_{\rm Lamb}$ in the $\ket{2}$ level:
\beq
\begin{aligned}
&\bra{2}H_\text{eff}\ket{2}\\
=&\sum_{k'}\frac{G(k')G^*(k')}{\Delta_2-2J\cos k'}\\
=&\frac{2g^2}{N}\sum_{k'}\frac{1}{\Delta_2-2J\cos k'}\mleft|\sum_{j=1}^{2}\exp(ik'x_j)\mright|^2\\
=&\frac{2g^2}{N}\sum_{k'}\frac{2+2\cos[k'(x_2-x_1)]}{\Delta_2-2J\cos k'} .
\end{aligned}
\eeq
For the parameters used in Fig.~3(a–d) of the main text, this Lamb shift takes a value of $\Delta_{\rm Lamb} \approx 0.057J$.

Finally, the doublon modes also get perturbed by $H_{int}$:
\beq
\begin{aligned}
&\bra{\psi_D(k_1)}H_\text{eff}\ket{\psi_D(k_2)}\\
=&\sum_{k'}\frac{G^*(k_1-k')G(k_2-k')}{\Delta_2-2J\cos k'}\\
=&\frac{2g^2}{N}\sum_{k'}\frac{e^{i(k_2-k_1)x_1}+e^{i(k_2-k_1)x_2}}{\Delta_2-2J\cos k'}\\
&+\frac{2g^2}{N}\sum_{k'}\frac{e^{i(k_2-k_1)x_1}e^{i(k_2-k')(x_2-x_1)}}{\Delta_2-2J\cos k'}\\
&+\frac{2g^2}{N}\sum_{k'}\frac{e^{i(k_2-k_1)x_2}e^{i(k_2-k')(x_1-x_2)}}{\Delta_2-2J\cos k'} ,
\end{aligned}
\eeq
which modifies both the doublon dispersion and the doublon eigenstates.

\begin{figure*}[htbp!]
    \centering
    \includegraphics[width=0.7\linewidth]{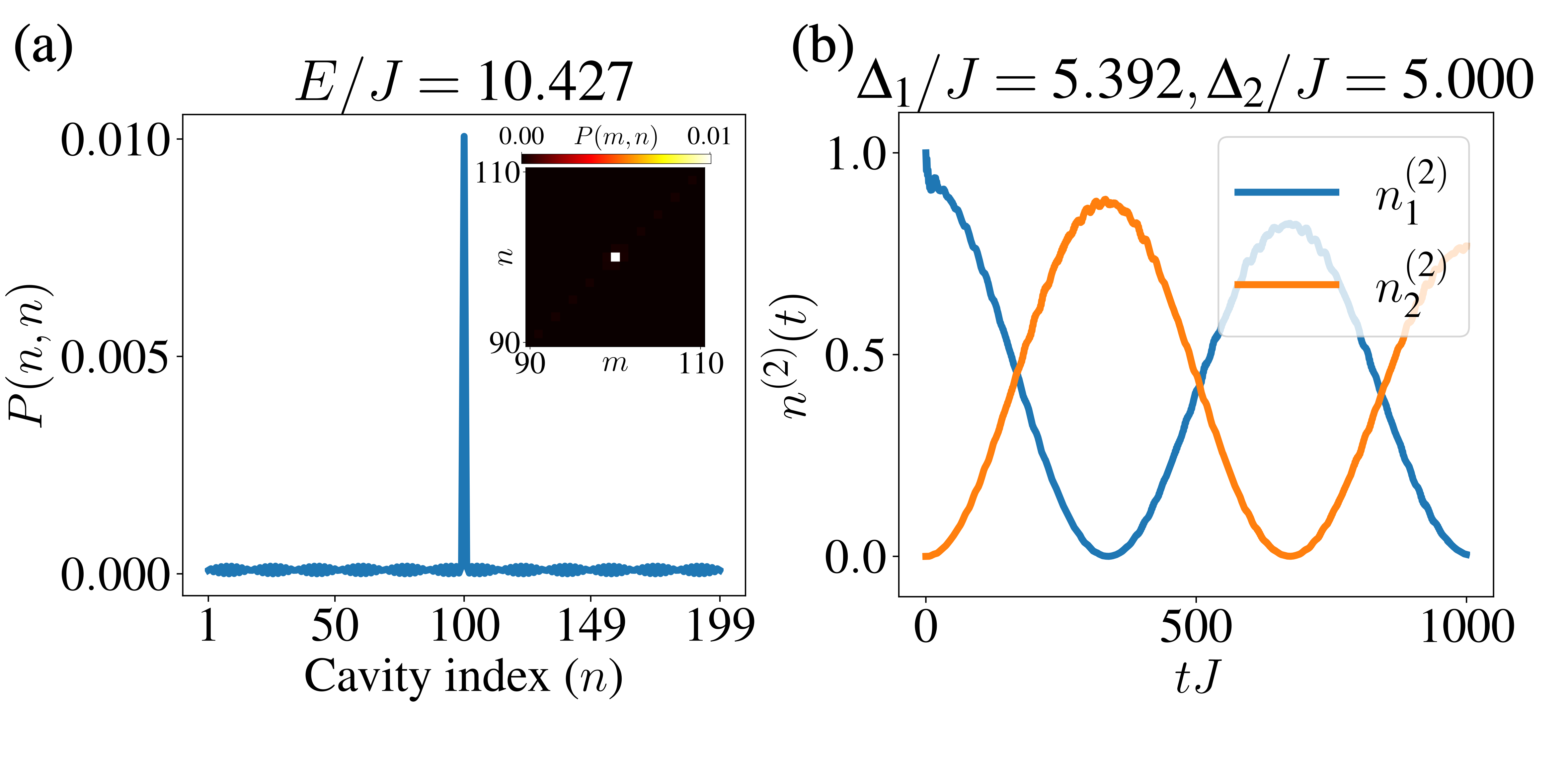}
    \caption{Doublon BIC and DFI without accounting for the Lamb shift of the $\ket{2}$ level for $U=10$, $g=0.25J$, $\Delta x=2$, $\Delta_1 = 5.392$ and $\Delta_2 = 5$. (a) The doublon BIC for a single giant atom, where an extension of the bound state wavefunction outside the coupling points ($99$ and $101$) can be seen, indicating that it is not a perfect bound state. 
    (b) The corresponding DFI dynamics for two giant atoms, where the dynamics show a gradual decay of the total atomic population, due to the imperfect BIC.}
    \label{figS1}
\end{figure*}

As a result, the decoherence-free condition also changes: satisfying $\Delta_1+\Delta_2=\omega_\text{DF}$ is no longer sufficient. The corrected condition becomes
\beq
\Delta_1+\Delta_2+\Delta_{\rm Lamb}=\omega_\text{DF}+\Delta_D(\omega_\text{DF})
\eeq
where $\Delta_D(\omega_\text{DF})$ is the perturbative shift in the doublon energy.

As shown in Fig.~\ref{figS1}, choosing $\Delta_1+\Delta_2=\omega_\text{DF}$ fails to produce a true BIC: the state energy is shifted, and the wavefunction spreads beyond the coupling region [\figpanel{figS1}{a}]. Correspondingly, the DFI quality deteriorates [\figpanel{figS1}{b}].

Numerically, we find that setting $\Delta_1+\Delta_2+\Delta=\omega_\text{DF}$ with $\Delta=0.054J$ results in the best decoherence-free quality for the parameters in Fig.~3 in the main text. The close match between $\Delta$ and $\Delta_{\rm Lamb}$ indicates that $H_{\rm int}$ only has a minor effect on the doublon dispersion.